\documentclass[aps,twocolumn,longbibliography,
reprint,%
]{revtex4-1}
\usepackage{amsmath}
\usepackage{amssymb}
\usepackage{graphicx,color,natbib,tabto}
\usepackage{dcolumn}
\usepackage{bm}
\usepackage{hyperref}
\hypersetup{
  colorlinks   = true, 
  urlcolor     = blue, 
  linkcolor    = blue, 
  citecolor   = blue 
}
\allowdisplaybreaks[1]

\def\bs#1{\boldsymbol{#1}}
\def\rmd{{\rm d}}
\def\rme{{\rm e}}
\def\rmi{{\rm i}}
\def\pdiff#1#2{{{\partial #1} \over {\partial #2}}}
\def\diff#1#2{{{\rmd #1} \over {\rmd#2}}}

\def\bs#1{\boldsymbol{#1}}

\begin{document}

\preprint{Preprint \#}

\title[Traveling waves in a continuum model for schooling swimmers]{Traveling waves in a continuum model of schooling swimmers}

\author{Anand U. Oza}%
 \email{oza@njit.edu}
\affiliation{ 
Department of Mathematical Sciences \& Center for Applied Mathematics and Statistics, New Jersey Institute of Technology, Cullimore Hall, Newark, NJ 07102, USA}%

\author{Eva Kanso}
 
\affiliation{ 
Department of Aerospace and Mechanical Engineering \& Department of Physics and Astronomy, University of Southern California, 3650 McClintock Avenue, Los Angeles, CA 90089, USA}%

\author{Michael J. Shelley}%
\email{mshelley@flatironinstitute.org}
\affiliation{
Courant Institute of Mathematical Sciences, 251 Mercer Street, New York, NY 10012, USA}
\affiliation{Center for Computational Biology, Flatiron Institute, 162 Fifth Avenue, New York, NY 10010, USA}%

\date{\today}

\begin{abstract}
 The complex formations exhibited by schooling fish have long been the object of fascination for biologists and physicists. However, the physical and sensory mechanisms leading to organized collective behavior remain elusive. On the physical side in particular, it is unknown how the flows generated by individual fish influence the collective patterns that emerge in large schools. To address this question, we 
 here present a continuum theory for a school of swimmers in an inline formation. The swimmers are modeled as flapping wings that interact through temporally nonlocal hydrodynamic forces, as arise when one swimmer moves through the lingering vortex wakes shed by the others, leading to a system of time-delay-differential equations. Through coarse-graining, we derive 
 a system of partial differential equations for the evolution of swimmer density and collective vorticity-induced hydrodynamic force. Linear stability analysis of the governing equations shows that there is a range of swimmer densities for which the uniform (constant-density) state is unstable to perturbations. Numerical simulations {in periodic domains} reveal families of stable traveling wave solutions, where a uniform school destabilizes into a collection of densely populated ``sub-schools" 
separated by relatively sparse regions that move as a propagating wave.
We find that distinct propagating waves may be stable for the same set of kinematic parameters. {We also find that finite schools can evolve into packets of coarsening traveling waves whose overall spreading is described by a rarefaction fan moving upstream and a terminating downstream shock.} Generally, our results show that temporally nonlocal hydrodynamic interactions can lead to rich collective behavior in schools of swimmers.
\end{abstract}

\keywords{Flocking, schooling, active matter, traveling waves}
\maketitle

\section{Introduction}

The complex dynamics of animal collectives such as fish schools offers a beautiful visualization of non-equilibrium collective behavior~\cite{Pavlov_Review}. A fish school is an archetype of an active matter system, as the individual organisms consume energy in order to self-propel and interact with each other~\cite{Marchetti_Review}. The mechanisms by which large groups of fish generate and sustain organized collective locomotion has long fascinated scientists~\cite{Shaw1962,Partridge1982}, one goal being to uncover their fundamental underlying biological principles and use them towards potential engineering applications, for example designing swarms of biomimetic underwater vehicles~\cite{Geder_2017}. 

Much of the prior research has focused on the behavioral mechanisms by which fish interact with each other while schooling~\cite{Parrish2002,LopezReview}. Experimental studies of different fish species in controlled environments have sought to uncover interaction rules that govern how fish move and orient themselves with respect to each other~\cite{Jolles1,Tunstrom1,Gautrais2012,Katz2011,HerbertRead2011,Yushi2022}. These studies are complemented by a variety of theoretical models that seek to capture the collective phenomena observed in biological systems. Discrete models, such as the seminal Viscek model~\cite{Viscek} and variants thereof (e.g.~\cite{Huth,Chate_Viscek,Calovi2015}), posit equations of motion for the coupled positional and orientational dynamics of interacting agents that sense each other and respond accordingly while being perturbed by stochastic noise. Continuum models have the advantage of being more amenable to analysis, and are biologically relevant when a collective contains a large number of individuals. For example, the celebrated Toner-Tu continuum partial differential equation (PDE) theory~\cite{Toner_1998,TonerTu_1995,Toner_2012} is a coarse-graining of the Viscek model, and has motivated other more sophisticated theories (e.g.~\cite{CavagnaPRL,Yang_Turning}). Partial integro-differential equations have been used to model spatially nonlocal sensory interactions~\cite{Mogilner_PDE,Topaz_2004,Carrillo_2013,Levertenz_2009}, wherein the constituents sense and thus interact with each other over long ranges. Despite their successes, these spatially nonlocal models do not account for temporally nonlocal interactions between constituents. Specifically, swimming fish generate flow structures that persist over spatial and temporal scales relevant to physical interactions within a school~\cite{Triantafyllou_Review}, yet modeling these interactions remains challenging due to the inherently non-local and non-instantaneous nature of these flows. 

A substantial body of research has demonstrated that hydrodynamic interactions are important in mediating schooling behavior~\cite{Ligman2023}. For example, studies have demonstrated that fish exert less energy while swimming in a school than they do in isolation~\cite{Marras,Herskin,Zhang2024_2}, especially in the turbulent flow conditions characteristic of relatively fast-moving schools~\cite{Zhang2024}. Hydrodynamic considerations have been used to rationalize the emergence of particular schooling formations over others in a water tank~\cite{AshrafPNAS}. In order to isolate hydrodynamic effects, recent experiments have been performed in which fish are replaced by freely-swimming wings, whose vertical (periodic) flapping motion is prescribed but horizontal motion is determined by the balance of hydrodynamic thrust and drag forces. These experiments have shown that a linear periodic array of flapping wings can move faster than a single wing~\cite{Becker}, and that pairs of wings in an in-line~\cite{Sophie} or staggered~\cite{Newbolt_2022,Ormonde2024} formation spontaneously adopt steady ``schooling modes," wherein the pair translates at constant velocity while maintaining a fixed separation distance. Such modes persist even if the wings' flapping amplitudes and frequencies are incommensurate~\cite{Newbolt2019}. Groups of more wings in an in-line formation exhibit flow-induced instabilities, wherein disturbances to the leader wing progressively amplify as they propagate downstream and can cause collisions between the trailing wings~\cite{Newbolt_2024}.

Theoretical models have been developed to describe hydrodynamic interactions between schooling swimmers in relatively high-Reynolds number flows. Models that account for far-field hydrodynamic effects, wherein each swimmer is modeled as a dipole~\cite{Kanso_Dipole,Gazzola_JFM}, have been used to demonstrate that swimmers adopt different schooling modes due to hydrodynamic interactions~\cite{Filella_Hydro}, and that swimming collectives undergo phase transitions when in confinement~\cite{Huang2024}. Near-field hydrodynamic interactions, which are due to the vortices generated by fish as they swim~\cite{Triantafyllou_Review}, have been modeled using a generalized thin-airfoil theory~\cite{Wu1961Swimmingwavingplate,Wu1975Extractionflowenergy}, wherein the periodic wake generated by a flapping plate is assumed to be flat~\cite{Sophie,Baddoo2022}. The vortices shed by fish have been modeled explicitly using point-vortex models, first in Refs.~\cite{Weihs1,Weihs2} and later generalized in Ref.~\cite{Oza2019}, which were used to compare the hydrodynamic thrust and efficiency of different schooling formations. Numerical simulations using a vortex sheet method have been used to assess the hydrodynamic interactions between pairs~\cite{Heydari_2021,Fang2025} and larger collectives~\cite{Heydari_2024,Nitsche_2025} of flapping plates in the limit of infinite Reynolds number.

However, there is currently no continuum description of swimmers that generate and interact through a relatively high-Reynolds number flow, a necessary step towards simulating the large fish schools observed in nature. The difficulty is that the hydrodynamic interactions between swimming fish are {\it temporally nonlocal}: while small-scale objects such as bacteria~\cite{2010SaintillanShelley} or cilia~\cite{Golestanian2011} generate a low-Reynolds number (Stokes) flow that can be readily incorporated into continuum models because the interactions between constituents are instantaneous, fish swim at higher Reynolds number and thus generate vortical flow structures that persist over a timescale $\tau$ that is often comparable to the timescale $d/u_0$ over which fish interact, $d$ and $u_0$ being the typical distance between and velocity of the fish, respectively. 

We here explicitly account for this temporal nonlocality by using a system of nonlinear delay-differential equations to model the {swimmers'} dynamics~\cite{Becker,Newbolt2019,Newbolt_2024}. The predictions of the model have been shown to exhibit favorable agreement with experiments on flapping wings in a water tank, and with both CFD and vortex sheet simulations of flapping wings~\cite{Heydari_2024}. While Viscek-type models with constant time-delay have been studied previously~\cite{Forgoston2008,Holubec_2021}, this model differs crucially in that the time-delays are state-dependent~\cite{Becker,Newbolt2019,Newbolt_2024}, which accounts for the fact that the swimmers interact with each other through a fluid-mediated memory. Another key feature of the model is that the interactions are non-reciprocal, in that a given swimmer is affected by the flow generated by the swimmers ahead of it. We make two simplifying assumptions: (1) we consider a kinematic model~\cite{Becker}, in which the swimmers' inertia is neglected so hydrodynamic forces directly determine the swimmers' velocities; and (2) we analyze the model in 1D, thus modeling a linear chain which is perhaps the simplest formation in which fish have been observed to school~\cite{Gudger}. By introducing new field variables that account for the persistent vortical flow generated by the swimmers, we show that this model coarse-grains naturally and thus obtain 
a 
continuum PDE description of schooling {swimmers} with temporally-nonlocal interactions. We show that, in certain parameter regimes, a uniform-density school of swimmers is unstable to perturbations. These perturbations give rise to traveling waves, which consist of densely populated ``sub-schools" of swimmers separated by relatively sparse regions. {Lastly, we numerically investigate the evolution of a compactly supported school.}

\section{Derivation of the continuum PDE model}\label{Sec:PDE}

\begin{figure}
    \includegraphics[width=.5\textwidth]{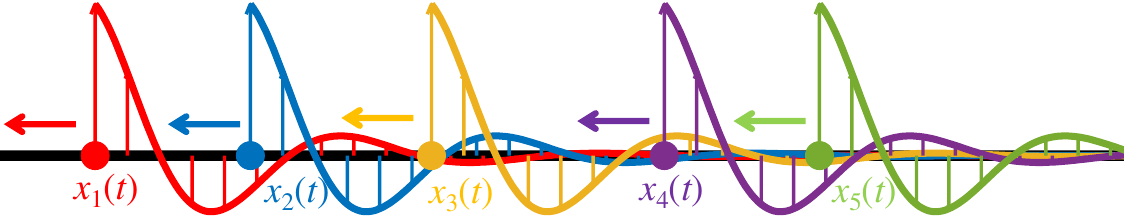}
      \caption{Schematic of the model~\eqref{Model2} for a linear school of flapping swimmers in a fluid. The swimmers are modeled as particles (dots), which are constrained to move in a line to the left. Each swimmer generates a wake whose vertical velocity (curves and arrows) oscillates and decays. These wakes induce horizontal forces on the downstream swimmers, in that a swimmer experiences a horizontal thrust (drag) if it moves up (down) into a downflow or down (up) into an upflow. The interactions between the swimmers are non-reciprocal, as each swimmer is influenced by the wake{s shed upstream of it.} 
      }
      \label{Schematic1}
\end{figure}

We present a derivation of the continuum PDE model, the details of which are given in Supplemental Material~\S\ref{App:Derivation}. We consider a collection of $N$ flapping swimmers with positions $x_i(t)$ that self-propel while maintaining an in-line formation (Fig.~\ref{Schematic1}). The swimmers are assumed to flap in-phase with the same flapping angular frequency $\omega$ and amplitude, and are assumed to move to the left, so $\dot{x}_i<0$. The time-evolution of their positions is governed by a system of nonlinear delay-differential equations, versions of which have been proposed previously to model the hydrodynamic interactions between flapping swimmers in high-Reynolds number flows~\cite{Becker,Newbolt2019,Newbolt_2024}. The equations are
\begin{subequations}\label{Model2}
\begin{align}
&\dot{x}_i=-u_0-{u_1}\sum_{j=1}^NK(t-t_j(x_i)),\label{Model2a} \\
&\text{where}\,\,{t_j(x)=\begin{cases} \text{arg}_t x_j(t)=x &\text{if }\exists t\text{ such that }x_j(t)=x, \\ +\infty &\text{otherwise};\end{cases}}\label{Model2b} \\
&K(t)=-H(t)\rme^{-t/\tau}\cos\omega t;\label{Model2c}
\end{align}
\end{subequations}
the constants $u_0$ and $u_1$ are positive; and $H(t)$ is the Heaviside step function. We have assumed that the swimmers' inertia is negligible and thus the hydrodynamic forces on the swimmers directly determine their velocities. 

What do these equations assume and encapsulate? In Eq.~\eqref{Model2a}, each swimmer is assumed to self-propel in isolation with a speed $u_0$, as determined by the balance between hydrodynamic thrust and drag forces on it. The swimmers interact through the term proportional to $u_1$, the physical interpretation being that each swimmer experiences hydrodynamic forces due to the vortical wakes {shed upstream of it.} 
Following Ref.~\cite{Newbolt2019}, we employ a simplified description of {the vortical} 
wakes: the combined wake is given by a linear superposition of the wakes generated by each swimmer, and the vertical wake speed generated by a given swimmer as it swims by is equal to its instantaneous vertical flapping velocity. This wake speed oscillates in time with frequency $\omega$ and decays exponentially over a timescale $\tau$ [Eq.~\eqref{Model2c}], the latter modeling the turbulent dissipation of flows at high Reynolds number~\cite{Higdon,Daghooghi,Sophie,Newbolt2019}. The force on swimmer $i$ is determined by its {own vertical velocity $\dot{y}_i$ relative to the local vertical velocity $v_i$ of the fluid flow generated by the other swimmers. Specifically, the force is proportional to $-\langle \left(\dot{y}_i-v_i\right)^2\rangle \approx T_0+T_1$, where $T_0=-\langle \dot{y}_i^2\rangle$, $T_1=2\langle \dot{y}_iv_i\rangle$ and $\langle\cdot\rangle$ denotes a time average over the flapping period, where we have assumed that $|v_i|\ll |\dot{y}_i|$. The free swimming speed $u_0$ is determined by the thrust on an isolated swimmer $T_0$, and the interaction force (proportional to $u_1$ in Eq.~\eqref{Model2a}) is determined by $T_1$ (Supplemental Material \S\ref{App:Derivation}). From the form of $T_1$, it is evident that a swimmer} experiences a thrust (drag) if it moves up (down) into a downflow or down (up) into an upflow. Equivalently, this force is determined by the swimmer's flapping phase relative to that of {the other swimmers} 
in the past. This relative phase is accounted for through the time-delays $t_j(x_i)$ defined in Eq.~\eqref{Model2b}, which specify the time in the past at which swimmer $j$ was in the current position of swimmer $i$. The non-reciprocal and causal nature of the interactions between swimmers is imposed by the Heaviside function $H(t)$, which ensures that a given swimmer is affected only by the 
{wakes generated} upstream of it. {That is, if swimmer $i$ remains ahead of swimmer $j$ for all time ($x_i(t) < x_j(t)$), it will never be affected by swimmer $j$, as $H\left(t-t_j(x_i(t)\right)=0$ 
and the corresponding interaction term in Eq.~\eqref{Model2a} would vanish. However, note that even if swimmer $i$ is ahead of swimmer $j$ at some instant $t$ ($x_i(t) < x_j(t)$), it would have been influenced by swimmer $j$ if it was behind swimmer $j$ at some earlier time $t^{\prime}<t$ ($x_i(t^{\prime})>x_j(t^{\prime})$).} Taken together, the vorticity-induced hydrodynamic interactions between flapping swimmers in high-Reynolds number flows are reduced to the non-reciprocal interactions of particles through memory: each particle leaves behind a signal, which oscillates and decays in time, and this signal is read by downstream particles as they pass by.

We convert the system of delay-differential equations~\eqref{Model2} into a system of ordinary differential equations by introducing the fields 
\begin{align}
C(x,t)&=\sum_{i=1}^NK(t-t_i(x))\nonumber \\
\text{and}\quad S(x,t)&=\sum_{i=1}^N\tilde{K}(t-t_i(x)),\nonumber \\
\text{where}\quad \tilde{K}(t)&=-H(t)\rme^{-t/\tau}\sin\omega t.
\end{align}
We also make the equations dimensionless via $t\rightarrow t/\tau$ and $x\rightarrow x/(u_0\tau)$, which introduces two dimensionless parameters: the strength of the hydrodynamic interactions $r = u_1/u_0$, and the dimensionless flapping frequency $\alpha=\omega\tau$. Note that
\begin{align}
\diff{}{t}H(t-t_i(x))=\delta(t-t_i(x))=\delta(x-x_i(t))|\dot{x}_i|,
\end{align}
where the second equality follows from the fact that $\delta(g(z))=\sum_i\delta(z-z_i)/|g^{\prime}(z_i)|$ for smooth functions $g(z)$ with roots at $z=z_i$, and the definition $t_i(x_i(t))=t$. Since $\dot{x}_i < 0$, we can rewrite Eq.~\eqref{Model2} as the system
\begin{subequations}\label{Micro1}
\begin{align}
\dot{x}_i&=-1-rC(x_i,t),\label{xi} \\
\partial_tC&=\sum_{i=1}^N\delta(x-x_i(t))\dot{x}_i-C-\alpha S,\label{Ci} \\ 
 \partial_tS&=-S+\alpha C.\label{Si}
\end{align}
\end{subequations}
The rewriting of Eq.~\eqref{Model2} as Eq.~\eqref{Micro1} is very convenient mathematically, as the temporally-nonlocal hydrodynamic interactions between swimmers (through the time delay functions $t_j(x)$) are replaced by spatial interactions through the aggregated field $C(x,t)$ generated by the swimmers. The oscillatory and decaying nature of the fields $C$ and $S$ (Fig.~\ref{Schematic1}) is evident through the relationship between them in Eqs.~\eqref{Ci}--\eqref{Si}.  {As an example, for a single swimmer ($N=1$) starting at the origin and moving at constant speed to the left, $x_1(t)=-Ut$, we have $C(x,t)=-\cos\left[\alpha(t+x/U)\right]\exp\left[-(t+x/U)\right]$ and $S(x,t)=-\sin\left[\alpha(t+x/U)\right]\exp\left[-(t+x/U)\right]$ if $-Ut<x<0$, and zero otherwise. That is, the fields are supported on the interval $[-Ut,0]$, where they exhibit decaying oscillations.} 

We now seek the {continuum} 
limit of Eq.~\eqref{Micro1}, since we expect that the empirical density 
${\sum_{i=1}^N\delta(x-x_i(t))}$ converges to a smooth density $\rho(x,t)$ as $N\rightarrow\infty$. We expect that $\rho$ satisfies the continuity equation $\partial_t\rho+\partial_x(\rho U_0)=0$~\cite{Golse2016,Jabin2017}, where we define the mean-field swimmer velocity $U_0(x,t)=-1-rC(x,t)$. Substituting Eq.~\eqref{xi} into Eq.~\eqref{Ci} and using the fact that $\delta(x-\tilde{x})g(\tilde{x})=\delta(x-\tilde{x})g(x)$, Eq.~\eqref{Ci} may be rewritten as
\begin{align}
\partial_t C&=
\rho U_0-{C}-\alpha{S}.
\end{align}
The resulting hyperbolic PDEs admit discontinuous solutions with shocks, so we regularize the equations by replacing $U_0\rightarrow U_{\nu}=U_0-\nu\partial_x(\log\rho)$, where $\nu > 0$ is an effective diffusivity. This additional term modifies the swimmers' velocities to decrease in the presence of a local increase in swimmer density. We thus obtain the equations
\begin{subequations}\label{PDE1}
\begin{align}
\partial_t\varrho&=\partial_x\left[(1+{C})\varrho\right]+\nu\partial_{xx}\varrho,\label{PDE1a} \\
 \partial_t{C}&=-\varrho\left(1+{C}\right)-\nu\partial_x\varrho-{C}-\alpha{S},\label{PDE1b} \\
 \partial_t{S}&=-{S}+\alpha{C},\label{PDE1c}
\end{align}
\end{subequations}
where we rescale the density, $\varrho(x,t)\equiv r\rho(x,t)$, and the fields $C\rightarrow rC$ and $S\rightarrow rS$. Note that the $\nu$--term leads to diffusive contributions in Eq.~\eqref{PDE1a}, which have been proposed in phenomenological PDE models for flocking and schooling~\cite{Toner_1998,TonerTu_1995,Yang_Turning}. While the swimmers in the biological setting under consideration (e.g. fish) are large enough that thermal fluctuations are insignificant, we hypothesize that random variations in the flapping frequencies, amplitudes or phases may be modeled by such diffusive terms. 


Equation~\eqref{PDE1} constitutes a system of three nonlinear PDEs for the fields $\varrho(x,t)$, $C(x,t)$ and $S(x,t)$. Equation~\eqref{PDE1a} describes the propagation of swimmer density $\varrho(x,t)$ through hydrodynamic interactions as induced by the field $C(x,t)$ and its diffusion. The field $C$ is generated by the swimmers through the terms proportional to $\varrho$ in Eq.~\eqref{PDE1b}, and is coupled to the field $S$ through Eq.~\eqref{PDE1c}. Through use of the field variables $C$ and $S$, we have analytically removed the delay terms in Eq.~\eqref{Model2}, which are otherwise difficult to handle analytically but are crucial for capturing the temporally nonlocal interactions between swimmers in relatively high-Reynolds number flows. 

\section{Linear instability of a uniform-density school}\label{Sec:LinStab}
The continuum model~\eqref{PDE1} admits a schooling state of uniform density, 
\begin{align}
\varrho=\varrho_0,\quad C=C_0\equiv -\frac{\varrho_0}{1+\alpha^2+\varrho_0},\quad S=S_0\equiv \alpha C_0.\label{UniformState}
\end{align}
The associated swimming speed $ 1+C_0$  is a monotonically decreasing function of $\varrho_0$, so the swimmers move slower as the density increases. A comparison between the steady states of the discrete [Eq.~\eqref{Model2}] and continuum [Eq.~\eqref{PDE1}] models is given in Supplemental Material \S\ref{Sec:ContDisc}. 

We study the linearized behavior of the system near this steady state. The linearized system admits plane-wave solutions, which we demonstrate directly by making the substitution $\varrho(x,t) = \varrho_0+\epsilon\hat{\varrho}(t)\rme^{\rmi kx}$, ${C}(x,t) = C_0+\epsilon\hat{C}(t)\rme^{\rmi kx}$ and ${S}(x,t) = S_0+\epsilon\hat{S}(t)\rme^{\rmi kx}$ into Eq.~\eqref{PDE1} and retaining terms at $O(\epsilon)$. We thus obtain a linear system of ODEs of the form $\dot{\bs{z}}=M\bs{z}$, where $\bs{z}=(\hat{\varrho},\hat{C},\hat{S})$ and
\begin{align}
M(k)=\begin{pmatrix} \rmi k(1+C_0)-\nu k^2 & \rmi k\varrho_0 & 0 \\ -(1+C_0)-\rmi k\nu & -(1+\varrho_0) & -\alpha \\ 0 & \alpha & -1\end{pmatrix}.\label{MMat}
\end{align}
Note that the linearized problem depends on the spatial wavenumber $k$, dimensionless flapping frequency $\alpha$, diffusivity $\nu$, and rescaled swimmer density $\varrho_0$.

\begin{figure*}
  \includegraphics[width=1\textwidth]{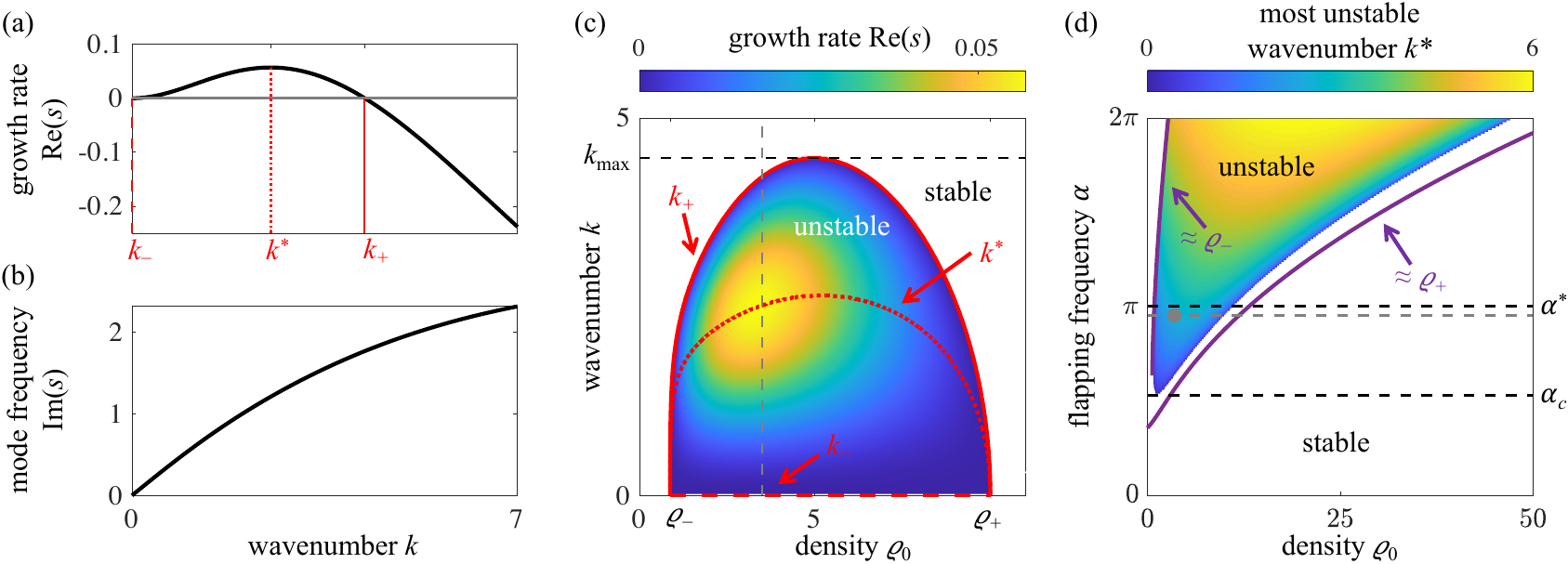}
      \caption{Results of the linear stability analysis of the uniform-density schooling state, as described in \S\ref{Sec:LinStab}. (a--b) Real (a) and imaginary (b) parts of the dominant eigenvalue $s(k)$ of the linear stability problem, for the dimensionless flapping frequency $\alpha=3$, rescaled density $\varrho_0 = 3.5$ and diffusivity $\nu=0.05$. The uniform state is unstable to perturbations with wavenumbers $k\in (k_-,k_+)$, with $k=k^*$ denoting the wavenumber with the largest growth rate. (c) Growth rate $\text{Re}(s)$ of the dominant eigenvalue, as a function of the school density $\varrho_0$ and perturbation wavenumber $k$, for $\alpha=3$ and $\nu = 0.05$. The stable region ($\text{Re}(s) < 0$) is indicated in white. The wavenumbers $k_-$, $k^*$ and $k_+$ are indicated by the dashed, dotted and solid red lines, respectively. The vertical gray line corresponds to the density $\varrho = 3.5$ used in panels (a)--(b). (d) Most unstable wavenumber $k^*$ as a function of $\alpha$ and $\varrho_0$, for fixed $\nu = 0.05$. The stable region is again indicated in white. The purple curves show the approximations~\eqref{rhoPM} for the critical densities $\varrho_{\pm}$. The black dashed lines correspond to the critical flapping frequencies $\alpha_c$ and $\alpha^*$. The gray dot indicates the $(\varrho_0,\alpha)$ pair corresponding to panels (a)--(b), and the dashed line denotes the value of $\alpha$ used in panel (c). The value of $\alpha$ used in panels (a)--(c) satisfies $\alpha_c(\nu)<\alpha<\alpha^*(\nu)$; this regime corresponds to scenario \#2 described in \S\ref{Sec:LinStab}, for which long-wave instabilities occur at $\varrho_0=\varrho_{\pm}$.}
      \label{StabFig1}
\end{figure*}

We find the three eigenvalues of $M(k)$ numerically, with eigenvalues with (positive) negative real part corresponding to linear (in)stability of the uniform-density schooling state~\eqref{UniformState}. Of the three eigenvalues, we restrict our attention to the one with the largest real part, which we call $s(k)$. The dependence of this eigenvalue on the wavenumber $k$, for fixed values of the parameters $\alpha$, $\varrho_0$ and $\nu$, is shown in Fig.~\ref{StabFig1}a--b. Our numerical computations of this eigenvalue for a range of parameters indicate that, when the uniform state is unstable, it is so for wavenumbers in a single finite interval, $k\in (k_-,k_+)$. Moreover, this eigenvalue has nonzero imaginary part, $\text{Im}(s) > 0$, suggesting that when the uniform-density state is unstable, it destabilizes into a dispersive traveling wave.

In Supplemental Material \S\ref{App:LinStab}, we analyze the eigenvalues of $M(k)$ in detail, in order to determine the parameter regimes in which the uniform-density schooling state~\eqref{UniformState} is unstable. Based on that analysis, we obtain the following results:
\begin{enumerate}
\item {\it Below a critical value of the flapping frequency, $\alpha < \alpha_c(\nu)$, all uniform-density schooling states are stable.}

That is, the uniform-density state can be unstable only if the dimensionless frequency $\alpha$ is sufficiently large [Supplemental Material~\ref{App:Long}]: 
\begin{align}
\alpha>\alpha_c(\nu)\equiv \sqrt{\frac{2}{1-4\nu(4/3)^3}-1}.\label{Crit1}
\end{align}
In particular, an instability exists only if the diffusivity is sufficiently small, $\nu < (3/4)^3/4\approx 0.1$, indicating that large levels of diffusion can eliminate the destabilizing influence of hydrodynamic interactions. 

\item {\it For $\alpha_c(\nu) < \alpha < \alpha^*(\nu)$, uniform-density schooling states with density $\varrho_0\in (\varrho_-,\varrho_+)$ are unstable. All such states satisfy $k_-=0$, as depicted in Fig.~\ref{StabFig1}c.}

That is, a long-wave instability occurs as the rescaled density is progressively increased or decreased from the stable regime, $\varrho_0\mathrel{\uparrow}\varrho_-$ or $\varrho_0\mathrel{\downarrow}\varrho_+$. Figure~\ref{StabFig1}c shows the dependence of the perturbation growth rate $\text{Re}(s)$ on the density $\varrho_0$ and perturbation wavenumber $k$ for a value of $\alpha$ in this regime. The critical densities $\varrho_{\pm}(\alpha,\nu)$, below and above which the uniform-density schooling state is stable, have the following asymptotic expansions in the limit $\nu\rightarrow 0$ [Supplemental Material~\ref{App:qLqR}]:
\begin{align}
\varrho_-(\alpha,\nu)&=\frac{\nu(1+\alpha^2)^2}{\alpha^2-1-4\nu(\alpha^2+1)}+O(\nu^3),\nonumber \\
\varrho_+(\alpha,\nu)&=(\alpha^2+1)\left[\left(\frac{\alpha^2-1}{\nu(\alpha^2+1)}\right)^{1/3}-\frac{4}{3}\right]\nonumber \\
&\phantom{=}+O(\nu^{1/3}).\label{rhoPM}
\end{align}
These asymptotic expressions for $\varrho_{\pm}(\alpha,\nu)$ are shown to exhibit good agreement with the numerically computed stability boundary in Fig.~\ref{StabFig1}d. The expressions in Eq.~\eqref{rhoPM} show that, if the rescaled density of swimmers is too low, $\varrho_0 < \varrho_-$, the hydrodynamic interactions are too weak to overcome the stabilizing influence of diffusion and generate an instability. Conversely, if the density is too large, $\varrho_0 > \varrho_+$, the diffusion-induced terms proportional to $\nu$ in Eq.~\eqref{PDE1b} cause the hydrodynamic interactions to destructively interfere and thus suppress the instability. Note that, in the limit $\nu\rightarrow 0$, the stability boundaries $\varrho_-$ and $\varrho_+$ shown in Fig.~\ref{StabFig1}d become vertical and horizontal lines, respectively, as the uniform-density state would be unstable for any $\varrho_0$ so long as $\alpha > 1$.

\item {\it For $\alpha > \alpha^*(\nu)$, uniform-density schooling states with density $\varrho_0\in (\varrho_-,\varrho_+)$ are unstable. Moreover, there is an intermediate density $\varrho^*(\nu)\in (\varrho_-,\varrho_+)$ above which $k_-=0$, as in \#2, but below which $k_->0$. 
}

That is, a {\it finite}-wavelength instability occurs at relatively low densities, as the density is progressively increased from the stable regime, $\varrho_0\mathrel{\uparrow} \varrho_-$; but a long-wave instability occurs at relatively high densities, as the density is progressively decreased from the stable regime, $\varrho_0 \mathrel{\downarrow}\varrho_+$ (Supplemental Fig.~\ref{fig:FiniteAmp}). As shown in Supplemental Material \S\ref{App:astar}, $\alpha^*(\nu)\rightarrow 1+\sqrt{2}$ as $\nu\rightarrow 0$.
\end{enumerate}

Figure~\ref{StabFig1}d shows the complete stability diagram for a fixed value of the diffusivity, $\nu = 0.05$, and offers a visualization of the foregoing results. Specifically, it shows the values of density $\varrho_0$ and flapping frequency $\alpha$ for which the uniform-density schooling state is stable (white) and unstable (colored). The colors indicate the most unstable perturbation wavenumber $k^*\in (k_-,k_+)$, defined as the wavenumber for which $\text{Re}(s)$ is the largest (Fig.~\ref{StabFig1}a). The fact that $\text{Im}[s(k^*)]\neq 0$ (Fig.~\ref{StabFig1}b) suggests that the schooling state undergoes a traveling wave instability with a critical length scale $\approx 2\pi/k^*$. We observe from Fig.~\ref{StabFig1}d that $k^*$ increases with $\alpha$, which indicates that a higher flapping frequency leads to waves with smaller wavelengths. While it is difficult to gain analytical insight into the dependence of $k^*$ on $\varrho_0$ and $\alpha$, 
in Supplemental Material \S\ref{App:kplus} we find a closed form expression for the upper bound $k_+$ of $k^*$ in the zero-diffusion regime ($\nu=0$):
\begin{align}
k_+(\varrho_0,\alpha)&= \frac{(1+\alpha^2+\varrho_0)(2+\varrho_0)}{2(1+\alpha^2)}\sqrt{\alpha^2-1}.\label{k_plus}
\end{align}
It follows from Eq.~\eqref{k_plus} that $k_+$ increases with $\alpha$ for $\nu=0$, in agreement with the trend for $k^*$ observed in Fig.~\ref{StabFig1}d. However, we note that the zero-diffusivity expression~\eqref{k_plus} predicts that $k_+$ increases monotonically with $\varrho_0$, which is at odds with the non-monotonic dependence of $k_+$ (and $k^*$) on $\varrho_0$ for $\nu > 0$ (Fig.~\ref{StabFig1}c). 
 
 The phase and group velocities of the most unstable mode ($k=k^*$) are negative with magnitudes less than unity for all values of $\alpha$ and $\varrho_0$, which indicates that the traveling waves propagate in the same direction as, but slower than, an isolated swimmer (Supplemental Fig.~\ref{fig:MiscStab}a--b). The phase velocity is larger in magnitude than the group velocity, as is typical for dispersive waves. Moreover, the magnitudes of the phase and group velocities  increase with the flapping frequency $\alpha$ but decrease with the density $\varrho_0$, which indicates that the waves move faster for larger flapping frequencies but slower for denser schools. The eigenvector $(\hat{\varrho},\hat{C},\hat{S})$ of $M(k^*)$, which is associated to the most unstable mode, has the property $\text{arg}(\hat{S})\approx \pi$ throughout the unstable regime (Supplemental Fig.~\ref{fig:MiscStab}c), where we fix $\hat{\varrho}=1$ without loss of generality. Physically, this means that oscillations in $\varrho$ are spatially out-of-phase with those of the field $S$, or that local increases in the swimmer density correspond to decreases in $S$. While the other component $\hat{C}$ varies more strongly with respect to $\varrho_0$ and $\alpha$ (Supplemental Fig.~\ref{fig:MiscStab}d), we observe that $\text{arg}(\hat{C})\in (-\pi,-\pi/2)$, indicating that the perturbation in $\varrho$ is related to that in $C$ by a phase shift of between a quarter- and half-wavelength.

\section{Traveling wave solutions in a periodic domain}\label{Sec:TWave}

\begin{figure*}
	\includegraphics[width=1\textwidth]{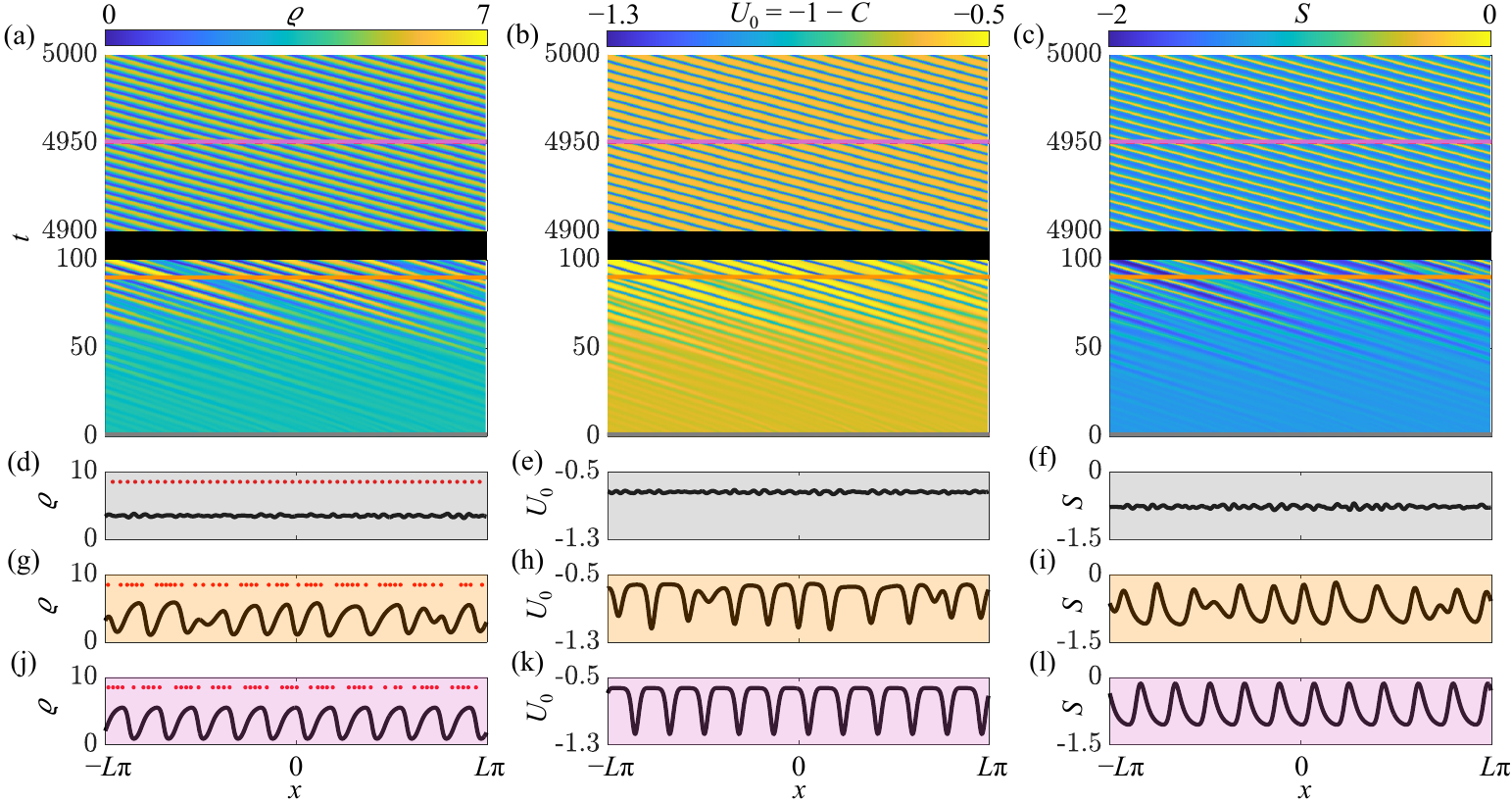}    
      \caption{Numerical simulation of Eq.~\eqref{PDE1} shows how small perturbations to a uniform-density schooling state amplify into a self-sustaining traveling wave. The parameters are the initial density $\varrho_0 = 3.5$, dimensionless flapping frequency $\alpha = 3$, diffusivity $\nu = 0.05$ and domain size $2\pi L$ where $L = 5$. (a--c) Spacetime plots of the density $\varrho$ (panel a) and fields $U_0= -1-C$ (panel b) and $S$ (panel c). (d--l) Snapshots of the solution at three distinct times $t$, indicated by the horizontal lines in panels (a--c) and color-coded appropriately. The density $\varrho$ is shown in panels (d,g,j), $U_0$ in (e,h,k) and $S$ in (f,i,l). Panels (d,g,j) also illustrate the evolving density using ``Lagrangian" particles that satisfy the ODEs $\dot{x}_i=U_0(x_i,t)$, with initial conditions $x_i(0)$ uniformly distributed on the interval shown. A movie of this solution is in Supplemental Movie 1.       }
      \label{Fig:Snapshots}
\end{figure*}

The linear stability analysis conducted in \S\ref{Sec:LinStab} suggests that the uniform-density schooling state~\eqref{UniformState} may destabilize through a traveling wave instability. We proceed by confirming the existence of traveling waves by presenting the results of numerical simulations of Eq.~\eqref{PDE1}. As described in Supplemental Material \S\ref{Sec:RungeKutta}, we simulate Eq.~\eqref{PDE1} on a periodic domain of length $2\pi L$ using a pseudospectral method combined with a fourth-order Runge-Kutta time-stepping scheme. We find that, if the initial data are chosen to correspond to the constant-density solution plus a small perturbation (Fig.~\ref{Fig:Snapshots}(d,e,f)), the fluctuations in $\varrho$, $C$ and $S$ grow quite rapidly in time (Fig.~\ref{Fig:Snapshots}(g,h,i)), before they saturate into a traveling wave solution of constant speed (Fig.~\ref{Fig:Snapshots}(j,k,l)). By comparing Fig.~\ref{Fig:Snapshots}(g,j) and Fig.~\ref{Fig:Snapshots}(i,l), we observe that $\varrho$ and $S$ are roughly out-of-phase with respect to each other, which is consistent with the linear stability theory in \S\ref{Sec:LinStab}. 

Supplemental Movie 1 and the spacetime plots in Fig.~\ref{Fig:Snapshots}(a,b,c) serve to further illustrate how the uniform-density school destabilizes into a traveling wave. 
The ``Lagrangian" particles shown in Fig.~\ref{Fig:Snapshots}(d,g,j) and Supplemental Movie 1 illustrate the mechanism by which the traveling wave emerges: an initially uniformly distributed collection of swimmers destabilizes into a collection of densely populated ``sub-schools" separated by relatively sparse regions. The leader of one of the sub-schools accelerates into the rear of the upstream sub-school, a process that continues periodically and thus generates a moving wave.
 

\begin{figure*}
    \includegraphics[width=1\textwidth]{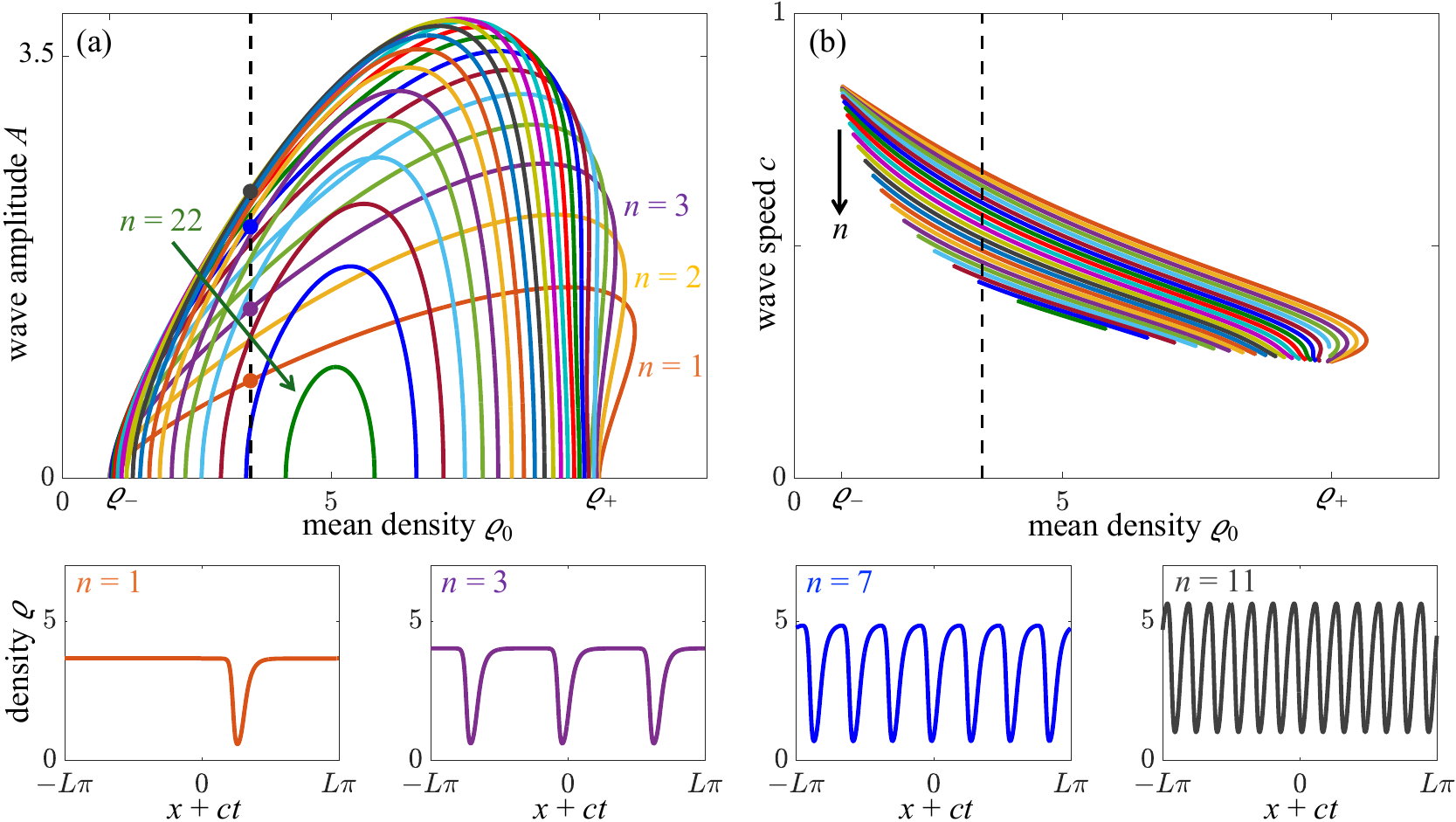}
      \caption{Traveling wave solutions to the continuum PDE model~\eqref{PDE1} for schooling swimmers, for the dimensionless flapping frequency $\alpha = 3$, diffusivity $\nu = 0.05$ and domain size $2\pi L$ where $L = 5$. The solutions are computed by solving the nonlinear eigenvalue problem~\eqref{TWave1} using the numerical procedure described in Supplemental Material \S\ref{App:TWave}. Panels (a) and (b) show the dependence of the wave amplitude $A$ and speed $c$ on the mean density $\varrho_0$, respectively. The branches are color-coded by their index $n$, which indicates the number of minima in the corresponding density profiles. The bottom panels show four density profiles for a fixed mean density $\varrho_0=3.5$, with the indicated values of $n$.}
      \label{fig:twave}
\end{figure*}

In order to systematically characterize the traveling wave solutions to Eq.~\eqref{PDE1}, we substitute the expressions $\varrho(x,t) = \varrho(x+ct)$, $C(x,t) = C(x+ct)$ and $S(x,t) = S(x+ct)$, where the wave profiles $\varrho$, $C$ and $S$ satisfy a system of nonlinear ODEs with periodic boundary conditions: 
\begin{align}
c\varrho^{\prime}-\left[(1+C)\varrho\right]^{\prime}-\nu \varrho^{\prime\prime}&=0,\nonumber \\
cC^{\prime}+\varrho(1+C)+\nu \varrho^{\prime}+C+\alpha S&=0,\nonumber \\
cS^{\prime}+S-\alpha C&=0.\label{TWave1}
\end{align}
Equation~\eqref{TWave1} constitutes a nonlinear eigenvalue problem in the wave speed $c$. We discretized the ODEs using a pseudospectral method, and then used MATLAB's root-finding algorithm to solve the resulting system of algebraic equations. Using the numerical method described in Supplemental Material \S\ref{App:TWave}, the traveling wave solutions were systematically tracked as a function of the density $\varrho_0 = \langle \varrho\rangle$ for fixed values of the flapping frequency $\alpha$ and diffusivity $\nu$, where $\langle \cdot\rangle$ denotes a spatial average. The solutions are characterized by their amplitude $A =\sqrt{2\langle \varrho^2\rangle}$ and wave speed $c$, which are plotted as functions of $\varrho_0$ in Fig.~\ref{fig:twave}(a,b). 

As shown in Fig.~\ref{fig:twave}a, for fixed values of $\alpha$ and $\nu$, a (finite) family of $n_{\mathrm{max}}=\lfloor k_{\text{max}}L\rfloor$ solution branches exists, where $k_{\text{max}} = \max_{\varrho_0}k_+(\varrho_0)$ is the largest unstable wavenumber across all densities $\varrho_0$ (Fig.~\ref{StabFig1}c). The branches may be indexed by the number of minima $n$ in the wave profiles $\varrho$, for $1\leq n\leq n_{\mathrm{max}}$ (Fig.~\ref{fig:twave}, bottom panels). Specifically, the $n=1$ branch corresponds to states of roughly constant density, punctuated by a single exponentially localized vacancy. The density profile of the vacancy is asymmetric about its minimum value, exhibiting a sharper gradient upstream (left) than downstream (right). As $n$ increases, the characteristic distance between high-density sub-schools decreases. The wave speed $c$ is less than unity (Fig.~\ref{fig:twave}b), indicating that the traveling waves move slower than a single isolated swimmer, which is consistent with the linear stability theory presented in \S\ref{Sec:LinStab} (Supplemental Fig.~\ref{fig:MiscStab}a--b). Similarly, the wave speed decreases with $n$, indicating that schools with a large number of density fluctuations move slower than relatively uniform ones. 
As the domain length $L\rightarrow\infty$, the diagram in Fig.~\ref{fig:twave}a will be populated by an infinite family of solutions, which we expect to effectively fill in the region bounded by the solution branches shown.

\begin{figure*}
	\includegraphics[width=1\textwidth]{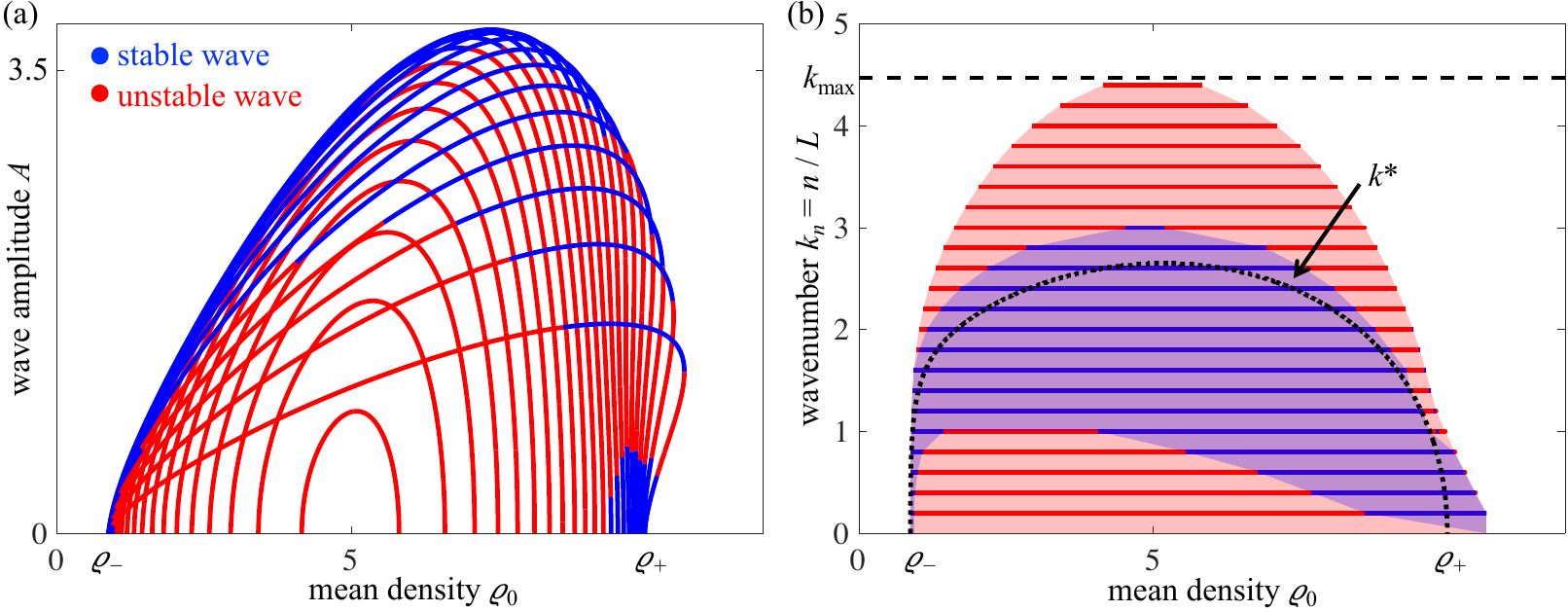}
      \caption{Stability of the traveling wave solutions shown in Fig.~\ref{fig:twave}. Blue (red) denote stable (unstable) solutions. (a) Traveling wave solutions shown in Fig.~\ref{fig:twave}a, now color-coded according to their stability. (b) Stability diagram of traveling wave solutions, obtained by taking the results in panel (a) and recasting them in terms of the solution's wavenumber $k_n\equiv n/L$. A point $(\varrho_0,k_n)$ is colored blue if at least one stable solution exists, and red otherwise. The black curve indicates the prediction of the most unstable wavenumber $k^*$ (Fig.~\ref{StabFig1}c) based on the linear stability analysis presented in \S\ref{Sec:LinStab}.}
      \label{Fig:TWaveStab}
\end{figure*}

Not all of the traveling wave solutions shown in Fig.~\ref{fig:twave} are stable. To assess their stability, we perform numerical simulations of the continuum PDE~\eqref{PDE1} up to the dimensionless time $t =5000$, with initial conditions given by the traveling waves shown in Fig.~\ref{fig:twave}, and determine whether the initial traveling wave persists. The results are shown in Fig.~\ref{Fig:TWaveStab}, with stable (unstable) solutions indicated in blue (red). Figure~\ref{Fig:TWaveStab}b shows that highly oscillatory traveling waves, with relatively large wavenumbers $k_n\equiv n/L$, are typically unstable. Low wavenumber waves are also typically unstable, except  at the highest swimmer densities $\varrho_0$. We observe that multiple stable traveling waves may coexist for the same density $\varrho_0$, indicating that linear schools may exhibit different stable wave modes.  The most unstable wavenumber $k^*$, as predicted by the linear stability analysis in \S\ref{Sec:LinStab} (Fig.~\ref{StabFig1}c), lies inside the region of stability for traveling waves (Fig.~\ref{Fig:TWaveStab}b). We conclude that, for a fixed density $\varrho_0$, the most (linearly) unstable wavenumber $k^*$ furnishes a reasonable prediction for the wavenumbers $k_n$ of the stable traveling waves that exist for that density. We also note that, for some of the branches at low $n$ (i.e. $n=1$), the amplitude $A$ is not a single-valued function of the density $\varrho_0$ for large $\varrho_0$. Indeed, some of these branches extend to the right of the linear stability boundary $\varrho=\varrho_+$ (Fig.~\ref{Fig:TWaveStab}a): while the corresponding uniform-density schooling states~\eqref{UniformState} are stable under infinitesimally small perturbations in this regime, they may destabilize under the influence of finite-amplitude perturbations.


We note that the value $\alpha=3$ considered in Fig.~\ref{fig:twave} and Fig.~\ref{Fig:TWaveStab} corresponds roughly to that found by Newbolt {\it et al.}~\cite{Newbolt2019} in their experiments on tandem flapping wings in a water tank. Specifically, they measured a decay time $\tau\approx 0.5$ s and considered wings with flapping frequency $f \approx 2$ Hz, for which $\alpha \approx 6$. We find qualitatively similar traveling wave solutions for a range of values of $\alpha$ and $\nu$.  

\section{Waves in a finite school}\label{Sec:Compact}

\begin{figure*} 
   \begin{center}
     \includegraphics[width=1\textwidth]{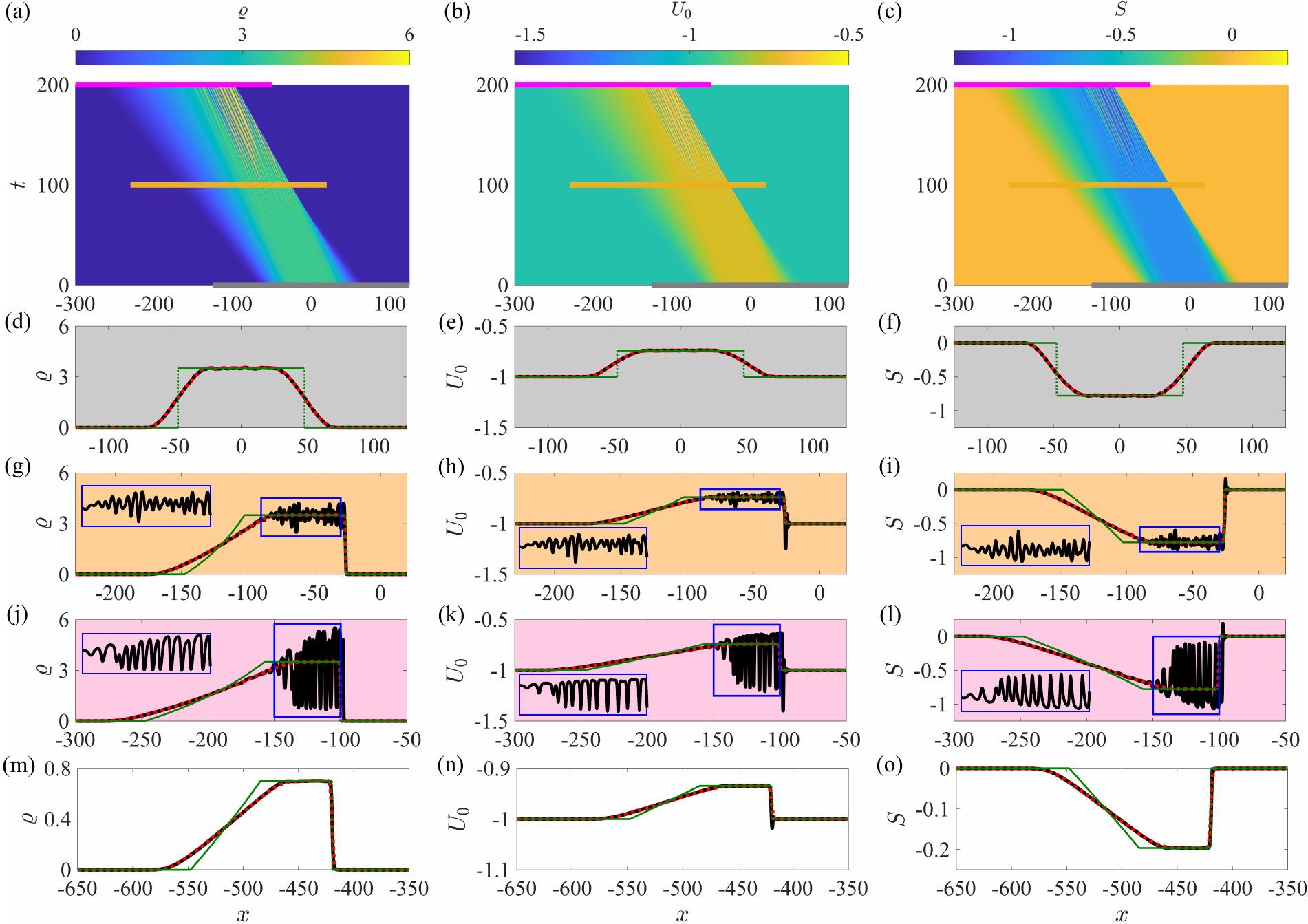}
      \end{center}
\caption{{Numerical simulation of the continuum model~\eqref{PDE1} with compactly supported initial data shows how a ``finite school" goes unstable via internal traveling waves. For panels (a)--(l), the parameters are $\varrho_0=3.5$, $\alpha=3$ and $\nu = 0.05$, as in Fig.~\ref{Fig:Snapshots}. Simulation details are provided in Supplemental Material \S\ref{App:Compact}. (a--c) Spacetime plots of the density $\varrho$ (panel a) and fields $U_0=-1-C$ (panel b) and $S$ (panel c). (d–l) Snapshots of the solution at three distinct times $t$, indicated by the horizontal lines in panels (a--c) and color-coded appropriately. The density $\varrho$ is shown in panels (d,g,j), $U_0$ in (e,h,k) and $S$ in (f,i,l).  In panels (g)--(l), the insets zoom into the interior region of the solution, where propagating waves are visible. The red dashed curves show the quasistatic solution, as obtained by solving Eq.~\eqref{PDEQS}. Specifically, the quasistatic density $\varrho_{\text{q}}$ is shown in panels (d,g,j), the associated velocity $U_{\text{q}}(\varrho_{\text{q}})$ in (e,h,k) and $S=-\alpha(1+U_{\text{q}}(\varrho_{\text{q}}))$ in (f,i,l). The green curves show the (discontinuous) solution to the quasistatic model with rectangular initial data and $\nu=0$, as given in Supplemental Material \S\ref{App:QSNuZero}. 
Panels (m)--(o) show a snapshot of the solution for a lower density, $\varrho_0 = 0.7$, at the time $t=500$. Note that the plots only show portions of the full solution in order to make the features more visible. Supplemental Movie 2 shows the time evolution of these two solutions.}
}
\label{Fig:Compact}
\end{figure*}

{The numerical solutions shown in Figs.~\ref{Fig:Snapshots} and~\ref{Fig:TWaveStab} were obtained in periodic domains for broadly supported initial data. To assess whether wave-like instabilities exist in a ``finite school," we conducted simulations of the continuum PDE~\eqref{PDE1} on a relatively large domain with initial data resembling a smooth flat-topped compactly supported bump function; details are given in Supplemental Material \S\ref{App:Compact}. A movie of the solution is shown in the top row of Supplemental Movie 2, and a spacetime plot of the simulation is shown in Fig.~\ref{Fig:Compact}(a,b,c).  Initially, the swimmer density $\varrho$ and associated fields $C$ and $S$ are roughly constant in a finite interval and vanish smoothly to zero outside a slightly larger interval, as shown in Fig.~\ref{Fig:Compact}(d,e,f).}

{We observe that the school goes unstable to density fluctuations, which amplify in a manner analogous to that observed in a periodic domain (Fig.~\ref{Fig:Snapshots} and Supplemental Movie 1). Specifically, small amplitude density fluctuations at early times grow within the interior of the school (Fig.~\ref{Fig:Compact}g), before saturating in amplitude and giving way to propagating waves. This traveling wave instability occurs when the swimmer density is still negligible at the edges of the simulation domain, indicating that it is not driven by the imposed periodicity of the solution.  While the rear of the school develops and sustains a sharp drop to nearly zero density, the front of the school progressively expands into the upstream region of low density. }

{It is clear from Supplemental Movie 2 (top row) that the solution has a relatively simple structure far upstream and downstream of the propagating waves in the interior. This structure may be described by a quasistatic model obtained by setting $\partial_tC=\partial_tS=0$ in Eqs.~\eqref{PDE1b}-\eqref{PDE1c} and solving for the associated fields $C=C_{\text{q}}$ and $S=\alpha C_{\text{q}}$, giving
\begin{align}
C_{\text{q}}=-\frac{\varrho_{\text{q}}+\nu\partial_x\varrho_{\text{q}}}{1+\alpha^2+\varrho_{\text{q}}}.
\end{align}
Substituting the expression for $C_{\text{q}}$ into Eq.~\eqref{PDE1a}, we obtain a single evolution equation for the quasistatic density $\varrho_{\text{q}}$:
\begin{align}
&\partial_t\varrho_{\text{q}}+\partial_x\left(U_{\text{q}}(\varrho_{\text{q}})\varrho_{\text{q}}\right)=\partial_x\left[D_{\text{q}}(\varrho_{\text{q}})\partial_x\varrho_{\text{q}}\right],\quad\text{where}\nonumber \\
&U_{\text{q}}(\varrho)=-1+\frac{\varrho}{1+\alpha^2+\varrho}
\text{ and } D_{\text{q}}(\varrho)=-\nu U_{\text{q}}(\varrho).\label{PDEQS}
\end{align}
This convection-diffusion equation is a generalization of the viscous Burgers' equation, with a nonlinear density-dependent correction on a leftwards wave moving at the base swimmer speed. The correction is positive and increases with $\varrho$, which indicates that the wave moves slower as the density is increased, a result that is consistent with the wave speeds obtained in the linear theory [Supplemental Fig.~\ref{fig:MiscStab}(a,b)] and in the periodic domain simulations (Fig.~\ref{fig:twave}b). The red dashed curves in Fig.~\ref{Fig:Compact}(d--l) and Supplemental Movie 2 show the quasistatic solution, obtained by solving Eq.~\eqref{PDEQS} numerically in a periodic domain using a method analogous to that presented in Supplemental Material \S\ref{Sec:RungeKutta}.  }

{An additional simplification may be obtained by considering the case of zero diffusion, $\nu=0$, and discontinuous ``rectangular" initial data, for which $\varrho_{\text{q}}(x,0)=\varrho_0$ for $|x| < x_0/2$ and $\varrho_{\text{q}}(x,0)=0$ otherwise, where $x_0$ is chosen such that the total mass $\varrho_0 x_0$ is the same as that of the smooth flat-topped bump function considered for the diffusive problem ($\nu> 0$). 
In this case, Eq.~\eqref{PDEQS} may be solved readily using the method of characteristics, as shown in Supplemental Material \S\ref{App:QSNuZero}. The solution is shown in the green curves in Fig.~\ref{Fig:Compact}(d--o) and Supplemental Movie 2, and has two salient features: first, a rarefaction wave of the form
\begin{align}
\varrho_{\text{q}}(x,t)&=(1+\alpha^2)\left(\sqrt{-\frac{t}{x+x_0/2}}-1\right)\label{RarefactionWave}
\end{align}
expands into an upstream region of zero density. Second, the solution has a discontinuity (or shock) at $x=x_{\text{s}}(t)$, where 
\begin{align}
x_{\text{s}}(t)=\begin{cases} \frac{x_0}{2}+U_{\text{q}}(\varrho_0)t&\text{if }t < t^*, \\ -\frac{x_0}{2}-\left(\sqrt{t}+\sqrt{\frac{(1+\alpha^2)x_0}{\varrho_0}}-\sqrt{t^*}\right)^2 &\text{if }t > t^*\end{cases}\label{ShockTraj}
\end{align}
is the shock trajectory and $ t^*\equiv x_0(1+\alpha^2+\varrho_0)^2/[(1+\alpha^2)\varrho_0]$ is the time at which the shock first reaches the downstream edge of the rarefaction wave. The density is positive upstream of the shock (at $x=x_{\text{s}}^-$) and zero downstream of it (for $x > x_{\text{s}}$). {The rarefaction wave expands relatively slowly for $t > t^*$, in that it is supported on the interval $(-t-x_0/2,x_{\mathrm{s}}(t))$ which is of length $\sim\sqrt{t}$ as $t\rightarrow\infty$.}}

{The agreement between the quasistatic solution, both for $\nu=0.05$ (red) and $\nu=0$ (green), and the solution to the full model~\eqref{PDE1} is excellent away from the interior region that contains the waves. It is important to note that, for the initial data corresponding to the solution shown in Fig.~\ref{Fig:Compact}, the region of relatively constant density has a value $\varrho\approx \varrho_0$ that is unstable by the linear theory presented in \S\ref{Sec:LinStab}. If the initial data is chosen so that $\varrho_0$ is linearly stable, the solution to the full model~\eqref{PDE1} is well-described by the quasistatic model~\eqref{PDEQS}, as shown in Fig.~\ref{Fig:Compact}(m,n,o) and the bottom row of Supplemental Movie 2.}

{In summary, the quasistatic model~\eqref{PDEQS} is able to capture the structure of the envelope of the solution to the full model~\eqref{PDE1}, which consists of a rarefaction wave that expands into a low-density upstream background and a downstream shock beyond which the density abruptly decreases. The rarefaction wave and shock trajectory are well-described by Eqs.~\eqref{RarefactionWave} and~\eqref{ShockTraj}, respectively, which were obtained by solving the quasistatic model for $\nu=0$. However, the quasistatic model is unable to capture the waves that develop and propagate within the envelope in the full model when the state $\varrho=\varrho_0$ is linearly unstable to perturbations. We note that, due to the expansion of the rarefaction wave, the swimmer density eventually becomes non-negligible at the edges of the simulation domain, by which time the front and rear of the school effectively interact with each other. Whether it is possible to find a compactly supported traveling wave solution that persists as $t\rightarrow\infty$ is beyond the scope of the present paper, {although it seems unlikely given the slow upstream spreading of the rarefaction wave}.}

\section{Discussion}\label{Sec:Discussion}

In this paper, we derived and analyzed 
a continuum model for schooling swimmers that accounts for interactions via a fluid-mediated memory. The dynamics of the swimmers is governed by a system of nonlinear delay-differential equations (Eq.~\eqref{Model2}), versions of which have previously been benchmarked against experimental~\cite{Becker,Newbolt2019,Newbolt_2024} and numerical~\cite{Heydari_2024} studies on schooling wings. The interactions between swimmers have two salient features: they are non-reciprocal, in that {if swimmer $i$ stays in front of swimmer $j$ for all time, swimmer $j$ experiences a force due to the wake generated by swimmer $i$, but swimmer $i$ is unaffected by swimmer $j$.} 
The interactions are also temporally nonlocal, in that they are determined by the relative phases of a given swimmer's flapping motion and that of the other swimmers in the past. By introducing the new field variables $C$ and $S$, which account for the oscillatory and decaying nature of these interactions, we coarse-grained the system analytically and obtained a 
continuum PDE theory~\eqref{PDE1} that we expect to be valid in the limit of a large number of swimmers, $N\rightarrow\infty$. 

Linear stability analysis of the PDE reveals that a uniform-density schooling state is unstable in an intermediate range of densities, $\varrho_0\in (\varrho_-,\varrho_+)$, provided that the dimensionless flapping frequency $\alpha$ is sufficiently large, $\alpha>\alpha_c$ (Fig.~\ref{StabFig1}). Both long-wave and finite-wavelength instabilities can occur for $\varrho_0= \varrho_-$, and long-wave instabilities occur for $\varrho_0= \varrho_+$ (Supplemental Material Fig.~\ref{fig:FiniteAmp}). The unstable states destabilize into traveling waves, wherein densely populated ``sub-schools” are separated by relatively sparse regions (Fig.~\ref{Fig:Snapshots}). A systematic characterization of these traveling waves {in periodic domains} reveals that they exist as distinct solution branches across a range of swimmer densities $\varrho_0$ (Fig.~\ref{fig:twave}a), with a given branch $n$ indicating the number of oscillations in the swimmer density profile. The traveling wave speed $c$ typically decreases with both the number of oscillations $n$ and the mean density $\varrho_0$ (Fig.~\ref{fig:twave}b). Numerical simulations of the PDE~\eqref{PDE1} reveal that, while not all of these traveling wave solutions are stable, multiple stable traveling wave solutions may coexist for identical values of the mean density $\varrho_0$, dimensionless flapping frequency $\alpha$ and diffusivity $\nu$ (Fig.~\ref{Fig:TWaveStab}). {Simulations of the PDE with compactly supported initial data show that a ``finite school", initially with roughly constant density in a finite interval, may go unstable to waves that propagate in the interior of an envelope that spreads due to a rarefaction wave (Fig.~\ref{Fig:Compact}).} {The interior waves initially increase in amplitude and subsequently coarsen, getting trapped within a slowly contracting interval.}

Traveling waves have been observed in many other models for self-propelled particles. For example, the Viscek model exhibits density waves~\cite{Gregoire_2004}, which arise through microphase separation between relatively low and high density regions~\cite{Solon_2015}. The density waves may coexist with the homogenous isotropic state~\cite{Caussin_2014}, which is qualitatively similar to the observation from our model that stable traveling waves may coexist with the uniform-density schooling state for $\varrho_0>\varrho_+$ (Fig.~\ref{Fig:TWaveStab}a). The traveling waves observed in our model also bear some resemblance to those exhibited in a model of chemotactic motility-induced phase separation (MIPS): unlike conventional MIPS where there is no directional bias~\cite{Cates2015}, systems that chemotax exhibit persistent directed motion in a given direction, and are observed to phase separate into bands that form traveling waves in certain parameter regimes~\cite{Zhao2023}. Similarly, the swimmers in our model move with a fixed velocity $-u_0$ in isolation, and the traveling waves exhibited by the collective can be interpreted as a sort of phase separation between dilute and dense phases. Traveling waves have also been observed in phenomenological flocking models that consist of partial integro-differential equations with both symmetric~\cite{Mogilner_PDE} and asymmetric~\cite{Milewski2008} interactions, the latter being more reminiscent of the non-reciprocal interactions encoded in our model~\eqref{Model2}. Wave-like behavior has also been observed in experimental~\cite{Newbolt_2024} and numerical~\cite{Heydari_2024,Nitsche_2025} studies of in-line formations of relatively small collectives of flapping {swimmers}, wherein disturbances to the leader {swimmer} propagate downstream and lead to collisions between the trailing {swimmers}.

Hydrodynamic interactions have been observed to induce traveling wave behavior in other biological active matter systems, for instance in microswimmers~\cite{Tsang2016,Tsang2018}, cilia~\cite{Golestanian2011,Wollin2011,Meng2021,Kanale2022,Chakrabarti2022} and algae~\cite{Brumley2012,Brumley2015} at low Reynolds numbers, and in nematode collectives at intermediate Reynolds numbers~\cite{Quillen2021,Peshkov2022}. However, our work is a theoretical demonstration of traveling wave behavior in a system governed by the high-Reynolds number hydrodynamic interactions characteristic of fish schools and bird flocks. Turning waves have been observed in bird flocks~\cite{Potts1984} and shimmering waves in fish schools~\cite{Pertzelan2023}, the latter being a visual manifestation of the escape waves exhibited by schooling fish in response to attack by a predator~\cite{HerbertRead2015,Poel2022}. While our model entirely neglects behavioral mechanisms, the hydrodynamics-induced traveling wave instability we have identified could complement other wave-like phenomena driven by behavior. {Moreover, the upstream spreading in our simulations could be mitigated in real schools if the swimmers employ a control mechanism, for instance by locally adjusting their speed.} To probe the interplay between hydrodynamics and behavior, one could extend the model to allow the swimmers to sense both each other~\cite{Filella_Hydro} and hydrodynamic pressure forces~\cite{Liao_Review,Leif_LateralLine}, as fish are thought to respond behaviorally to such cues in natural settings. 
Generalizations of the theory in which the swimmers move in three dimensions while changing their swimming direction might be used to model collectives with more complex internal structure.

The main predictions of our continuum theory are that schooling swimmers in a linear formation exhibit traveling waves whose speed decreases both with the mean density and the number of density oscillations, which could be tested in experimental or observational studies of fish schools. Moreover, we presume that sufficiently strong external perturbations to a fish school could induce switching between the different multistable traveling wave modes, an effect that could be important in rationalizing  the complex dynamics exhibited by fish schools. Incorporating the swimmers' inertia into the model could lead to modulational and other more complex instabilities. Inertial effects are thought to be relevant in understanding the dynamics of bird flocks~\cite{AttanasiNatPhys,Cavagna2015,Cavagna2017}, and have been shown to significantly influence the dynamics of certain active matter systems~\cite{Lowen2020}. These worthwhile future directions would build on the modeling and analytical techniques presented herein, which provide a general framework for incorporating fluid-mediated memory into a continuum theory for active matter systems.

\acknowledgments{We thank Leif Ristroph and Miles Wheeler for useful discussions. A.U.O. acknowledges support from NSF DMS-2108839 and NSF DMS-2510304. Funding to E.K. is provided by the NSF grants RAISE IOS-2034043 and CBET-210020 and the Office of Naval Research grants N00014-22-1-2655 and N00014-19-1-2035.}

\bibliography{/Users/Anand/Dropbox/bibFiles/SchoolingBib.bib,/Users/Anand/Dropbox/bibFiles/ALCBib.bib}
\bibliographystyle{unsrtabbrv} 

\clearpage

\onecolumngrid
\setcounter{page}{1}
\setcounter{section}{0}
\setcounter{figure}{0}
\renewcommand{\thefigure}{S\arabic{figure}}
\renewcommand{\theequation}{S\arabic{equation}}
\setcounter{equation}{0}
\linespread{1}
\begin{center}
\large
{\bf SUPPLEMENTARY MATERIAL \\ ``Traveling waves in a continuum model for schooling swimmers"}
\end{center}
\normalsize

\section{Derivation of the continuum PDE}\label{App:Derivation}

We here present the details of the derivation of the continuum PDE given in \S\ref{Sec:PDE} of the Main Text. Our starting point is a discrete particle model that is based on the one in Ref.~\cite[Supplemental Material]{Newbolt2019}, proposed therein for $N=2$ {swimmers}. Assuming that the {swimmers} have identical masses $m$ and flap with the same amplitude and frequency $\omega$ with no phase offset, the horizontal positions of the {swimmers} evolve according to the equations
\begin{align}
m\ddot{x}_1+F_{\text{D}}(\dot{x}_1)&=-F_0,\nonumber \\
m\ddot{x}_2+F_{\text{D}}(\dot{x}_2)&=-F_0+F_1\cos\left[\omega(t-t_2(x_1))\right]\rme^{-(t-t_2(x_1))/\tau}-F_2\rme^{-2(t-t_2(x_1))/\tau},\label{DiscreteModel1}
\end{align}
where $F_{\text{D}}(\dot{x})$ is the drag force, $F_0>0$ is the thrust generated by an isolated {swimmer}, and $F_1,F_2>0$ are constants that influence the force on the follower ($x_2$) due to the wake generated by the leader ($x_1$). In deriving this model, it is assumed that the vertical velocity of the leader's wake is equal to the leader's vertical flapping speed as it swims by. This wake speed is then assumed to decay exponentially in time as an approximation of viscous and turbulent dissipation~\cite{Higdon,Daghooghi,Sophie,Newbolt2019}. It is also assumed that the thrust on a {swimmer} is proportional to the square of its vertical velocity component relative to the ambient fluid. Time-averaging these thrust forces over the flapping period leads to the terms on the right-hand side of Eq.~\eqref{DiscreteModel1}.

The generalization of Eq.~\eqref{DiscreteModel1} to $N$ {swimmers} in an in-line formation is
\begin{align}
m\ddot{x}_i+F_{\text{D}}(\dot{x}_i)&=-F_0+F_1\sum_{j=1}^NK(t-t_j(x_i)),
\end{align}
where we have removed the $F_2$-term in Eq.~\eqref{DiscreteModel1} because it is exponentially smaller than the $F_1$-term. We now let $x_i(t)=-u_0t+\tilde{x}_i(t)$ and retain terms at leading order in $\tilde{x}_i$; that is, we seek trajectories that exhibit small fluctuations around the steadily-translating state. The resulting equations are
\begin{align}
\frac{m}{F_{\text{D}}^{\prime}(-u_0)}\ddot{x}_i+\dot{x}_i&=-u_0+\frac{F_1}{F_{\text{D}}^{\prime}(-u_0)}\sum_{j=1}^NK(t-t_j(x_i)).
\end{align}
We non-dimensionalize these equations using $x\rightarrow x/(u_0\tau)$ and $t\rightarrow t/\tau$, as in \S\ref{Sec:PDE} of the Main Text. The non-dimensional equations are thus
\begin{align}
\frac{m}{F_{\text{D}}^{\prime}(-u_0)\tau}\ddot{x}_i+\dot{x}_i&=-1+\frac{F_1}{F_{\text{D}}^{\prime}(-u_0)u_0}\sum_{j=1}^NK(t-t_j(x_i)).
\end{align}
We now {assume that} inertia can be neglected because $m/(F_{\text{D}}^{\prime}(-u_0)\tau)\ll 1$,  
which yields the dimensionless version of Eq.~\eqref{Model2} in the Main Text.

As shown in \S\ref{Sec:PDE} of the Main Text, Eq.~\eqref{Model2} can be rewritten as Eq.~\eqref{Micro1}. We proceed by writing Eq.~\eqref{Micro1} in the form
\begin{align}
\dot{x}_i=-1-rC(x_i,t),\quad C(x,t)=\sum_{i=1}^N\int_0^t\mathrm{d}s\,\delta(x-x_i(s))\dot{x}_i(s)\cos\left[\alpha(t-s)\right]\rme^{-(t-s)}.
\end{align}
To obtain the {continuum} limit of this model, we consider the regularized model
\begin{align}
\dot{x}_{i,\epsilon}=-1-rC_{\epsilon}(x_{i,\epsilon},t),\quad C_{\epsilon}(x,t)=\sum_{i=1}^N\int_0^t\mathrm{d}s\,\delta_{\epsilon}(x-x_{i,\epsilon}(s))\dot{x}_{i,\epsilon}(s)\cos\left[\alpha(t-s)\right]\rme^{-(t-s)},
\end{align}
where $\delta_{\epsilon}(x)\rightarrow\delta(x)$ in the sense of distributions as $\epsilon\rightarrow 0$. We have
\begin{align}
C_{\epsilon}(x,t)&=\sum_{i=1}^N\int_0^t\mathrm{d}s\,\delta_{\epsilon}(x-x_{i,\epsilon}(s))\left[-1-rC_{\epsilon}(x_{i,\epsilon}(s),s)\right]\cos\left[\alpha(t-s)\right]\rme^{-(t-s)}\nonumber \\
&=\sum_{i=1}^N\int_0^t\mathrm{d}s\int_I\mathrm{d}\xi\,\delta(\xi-x_{i,\epsilon}(s))\delta_{\epsilon}(x-\xi)\left[-1-rC_{\epsilon}(\xi,s)\right]\cos\left[\alpha(t-s)\right]\rme^{-(t-s)}
\end{align}
where $I$ is the spatial ($x$) domain over which the equations are defined. In the {limit} $N\rightarrow\infty$, we assume that the empirical {density} $\sum_{i=1}^N\delta(x-x_{i,\epsilon}(t))$ converges to {a smooth function} $\rho_{\epsilon}(x,t)$, which satisfies the McKean-Vlasov PDE~\cite{Golse2016,Jabin2017}
\begin{align}
&\partial_t\rho_{\epsilon}-\partial_x\left[(1+rC_{\epsilon})\rho_{\epsilon}\right]=0,\nonumber \\
&C_{\epsilon}(x,t)=\int_0^t\mathrm{d}s\int_I\mathrm{d}\xi\,\delta_{\epsilon}(x-\xi)
\rho_{\epsilon}(\xi,s)\left[-1-rC_{\epsilon}(\xi,s)\right]\cos[\alpha(t-s)]\rme^{-(t-s)}.
\end{align}
We now take the limit $\epsilon\rightarrow 0$. Assuming that the limit can be taken inside the integral, and that $\rho_{\epsilon}$ and $C_{\epsilon}$ converge to functions $\rho$ and $C$, we obtain
\begin{align}
&\partial_t\rho-\partial_x\left[(1+rC)\rho\right]=0,\nonumber \\
&C(x,t)=\int_0^t\mathrm{d}s\,
\rho(x,s)\left[-1-rC(x,s)\right]\cos[\alpha(t-s)]\rme^{-(t-s)},
\end{align}
which is equivalent to Eq.~\eqref{PDE1} in the Main Text for $\nu=0$.

\section{Uniform density solutions in the discrete particle and continuum models}\label{Sec:ContDisc}

\begin{figure}[ht]
   \begin{center}
    \includegraphics[width=1\textwidth]{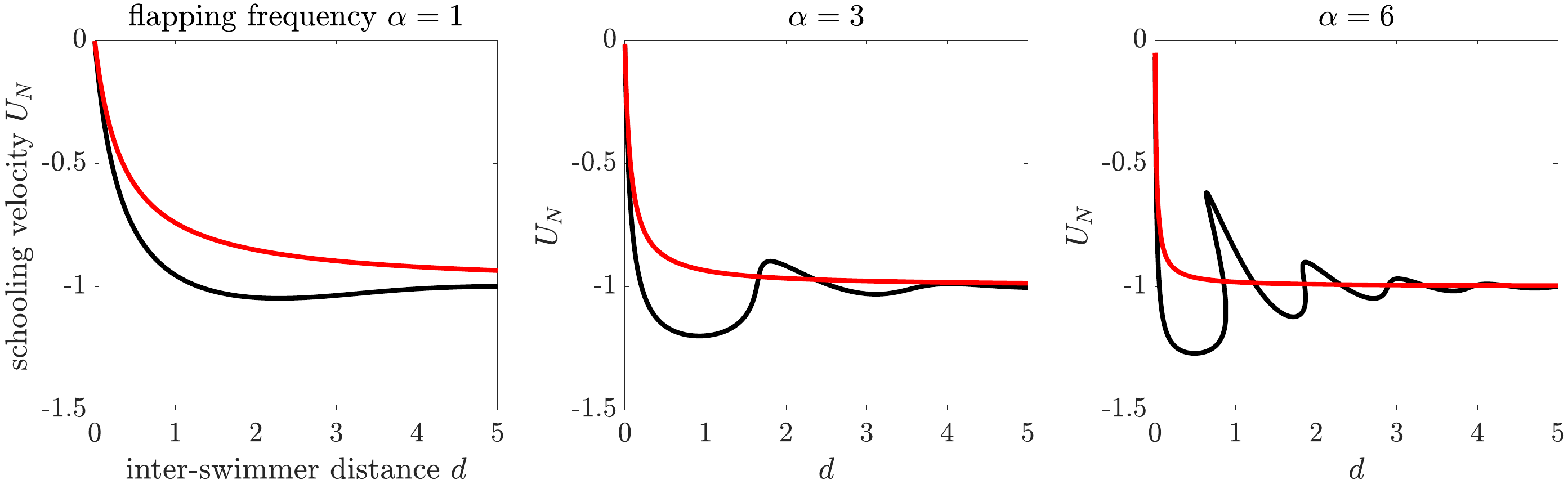}
    \end{center}
\caption{Comparison between the steady-state solutions to the discrete particle model (black, Eq.~\eqref{ParticleBS1}) and the continuum model (red, Eq.~\eqref{ContVel1}), for ${r}=0.7$ and dimensionless flapping frequency $\alpha=1$ (left), $\alpha=3$ (middle) and $\alpha=6$ (right).}
\label{fig:Particle1}
\end{figure}

We here deduce the relationship between uniform-density solutions to the discrete particle model and the continuum model, given in Eqs.~\eqref{Model2} and~\eqref{PDE1}, respectively, in the Main Text. To that end, we first note that the constant density state $\varrho=\varrho_0$ has velocity
\begin{align}
U_0\equiv -1-C_0 &= 
-\frac{1+\alpha^2}{1+\alpha^2+\varrho_0}\label{ContVel1}.
\end{align}
The discrete particle model in Eq.~\eqref{Model2} in the Main Text admits solutions in which $N$ swimmers on a periodic domain of length $\ell$ are equally spaced by a distance $d=\ell/N$ and move at a speed $U_N < 0$. In dimensionless variables, $U_N$ satisfies the algebraic equation
\begin{align}
U_N &= -1+{r}\sum_{n=1}^{\infty}\rme^{nd/U_N}\cos\frac{n\alpha d}{U_N}=
-1+{r}\left(\frac{\rme^{d/U_N}\cos\frac{\alpha d}{U_N}-\rme^{2d/U_N}}{1+\rme^{2d/U_N}-2\rme^{d/U_N}\cos\frac{\alpha d}{U_N}}\right).\label{ParticleBS1}
\end{align}
Equation~\eqref{ParticleBS1} is an implicit equation for $U_N$, which may have multiple solutions for a given $d$, as shown in Fig.~\ref{fig:Particle1} (black curves). We note that such a multiplicity of solutions was demonstrated in prior work~\cite{Becker}.

 In the {limit $d\rightarrow 0$ ($N\rightarrow\infty$)}, Eq.~\eqref{ParticleBS1} {may be expanded as}
\begin{align}
U_{\infty} = -1-\frac{rU_{\infty}}{{d}(1+\alpha^2)}\quad\Rightarrow\quad U_{\infty} = -\frac{1+\alpha^2}{1+\alpha^2+r/{d}}.
\end{align}
Note that $U_{\infty}$ is equal to the mean-field velocity~\eqref{ContVel1} in the continuum PDE since $\varrho_0=r/{d}$. The steady state of the discrete particle model~\eqref{Model2} thus  converges to that of the continuum PDE~\eqref{PDE1} {in the joint limit $r\rightarrow 0$ and $N\rightarrow\infty$ with $rN\rightarrow$ constant, which is the limit of weak interactions and high swimmer density.} However, while the discrete particle model may have multiple steadily-translating solutions for a given (finite) $N$, as demonstrated by the black curves in Fig.~\ref{fig:Particle1}, the continuum PDE has a unique steadily-translating solution for a given uniform density $\varrho_0$ (red curves in Fig.~\ref{fig:Particle1}).

\section{Linear stability analysis of the uniform density state}\label{App:LinStab}

We here detail the linear stability analysis of the uniform density state, the results of which were summarized in \S\ref{Sec:LinStab} of the Main Text. The characteristic polynomial $F(s;k)=\text{det}(M(k)-sI)$ of the matrix $M(k)\in\mathbb{C}^{3\times 3}$ defined in Eq.~\eqref{MMat} of the Main Text is $F(s;k) = s^3+f_2(k)s^2+f_1(k)s+f_0(k)$, where
\begin{align}
f_0(k)&=(1+\alpha^2)\left(\nu k^2 -\rmi k\frac{1+\alpha^2}{1+\alpha^2+\varrho_0}\right),\quad f_1(k)=1+\alpha^2+\varrho_0+2\nu k^2-2\rmi k\frac{1+\alpha^2}{1+\alpha^2+\varrho_0},\nonumber \\
f_2(k)&=2+\varrho_0+\nu k^2-\rmi k\frac{1+\alpha^2}{1+\alpha^2+\varrho_0}.
\label{fiFuncs}
\end{align}
The eigenvalues $s(k)$ of the linear stability problem are the roots of $F(s;k)$, with $\text{Re}(s) < 0$ ($ > 0$) indicating linear (in)stability. We label the eigenvalues by their real parts at $k=0$, $\text{Re}(s_3(0)) < \text{Re}(s_2(0)) < \text{Re}(s_1(0))$. Our numerical computations have shown that $s_1$ is dominant, in the sense that $\text{Re}(s_3(k)) < \text{Re}(s_2(k)) < \text{Re}(s_1(k))$ for all $k$, and that $\text{Re}(s_2) < 0$ and $\text{Re}(s_3) < 0$. We thus refer to $s_1(k)$ as $s(k)$ in the Main Text, as it determines whether the uniform-density state is linearly stable or unstable. 

We proceed by assessing the asymptotic behavior of the eigenvalues in the limits $k\rightarrow 0$ and $k\rightarrow\infty$ (\S\ref{App:Asymp}) and delineating the parameter regime in which the uniform state is unstable (\S\ref{App:Instab}). The special cases of long-wave instabilities ($k\rightarrow 0$) and zero diffusion ($\nu=0$)   and are treated in \S\ref{App:Long} and \S\ref{App:kplus}, respectively.

\subsection{Asymptotic formulas for the eigenvalues}\label{App:Asymp}

We first determine the long-wave behavior of the eigenvalues, and so compute $F(s,k)$ in the limit $k\rightarrow 0$:
\begin{align}
\lim_{k\rightarrow 0}F(s;k)=s\left(s^2+(2+\varrho_0)s+1+\alpha^2+\varrho_0\right).
\end{align}
The two nontrivial roots of this polynomial are $s=-1-\varrho_0/2\pm\sqrt{(\varrho_0/2)^2-\alpha^2}$. The asymptotic behavior of the third eigenvalue is deduced by inserting the expansion $s(k)=\tilde{s}k+O(k^2)$ into the expression for $F(s;k)$:
\begin{align}
F(s(k);k)=\left[(1+\alpha^2+\varrho_0)\tilde{s}-\rmi\frac{(1+\alpha^2)^2}{1+\alpha^2+\varrho_0}\right]k+O(k^2).
\end{align}
We thus obtain the solution $\tilde{s}=\rmi\left[(1+\alpha^2)/(1+\alpha^2+\varrho_0)\right]^2$. In summary, we have the following long-wave behavior:
\begin{align}
s_1 &= \rmi\left(\frac{1+\alpha^2}{1+\alpha^2+\varrho_0}\right)^2k+O(k^2),\quad
s_2 = -1-\frac{\varrho_0}{2}+\sqrt{\left(\frac{\varrho_0}{2}\right)^2-\alpha^2}+O(k),\nonumber \\
\quad\text{and}\quad s_3 &= -1-\frac{\varrho_0}{2}-\sqrt{\left(\frac{\varrho_0}{2}\right)^2-\alpha^2}+O(k)\quad\text{as}\quad k\rightarrow 0.\label{EigLong}
\end{align}

To determine the short-wave behavior of the eigenvalues, we first consider the behavior of $F$ as $k\rightarrow\infty$:
\begin{align}
\lim_{k\rightarrow\infty}\frac{F(s;k)}{k^2}=\nu\left(s^2+2s+1+\alpha^2\right),
\end{align}
which yields the two roots $s = -1\pm\alpha\rmi$ associated with $s_1$ and $s_2$. The asymptotic behavior of $s_3$ is deduced by assuming that $s(k)\sim\hat{s}k^2$ as $k\rightarrow\infty$:
\begin{align}
\lim_{k\rightarrow\infty}\frac{F(s(k);k)}{k^6}=\hat{s}^2\left(\hat{s}+\nu\right),
\end{align}
which yields the solution $\hat{s}=-\nu$. In summary, we have the following short-wave behavior:
\begin{align}
s_1\rightarrow -1+\alpha \rmi,\quad s_2\rightarrow -1-\alpha \rmi,\quad \text{and}\quad s_3\rightarrow -\nu k^2\quad\text{as}\quad k\rightarrow\infty.\label{EigShort}
\end{align}
The asymptotic predictions for $s_1$ in Eqs.~\eqref{EigLong} and~\eqref{EigShort} agree well with the numerically computed eigenvalue shown in Fig.~\ref{StabFig1}a of the Main Text. We also find favorable agreement for $s_2$ and $s_3$ (not shown).

\subsection{Parameter regime corresponding to linear instability}\label{App:Instab}

\begin{figure}[ht]
   \begin{center}
    \includegraphics[width=0.5\textwidth]{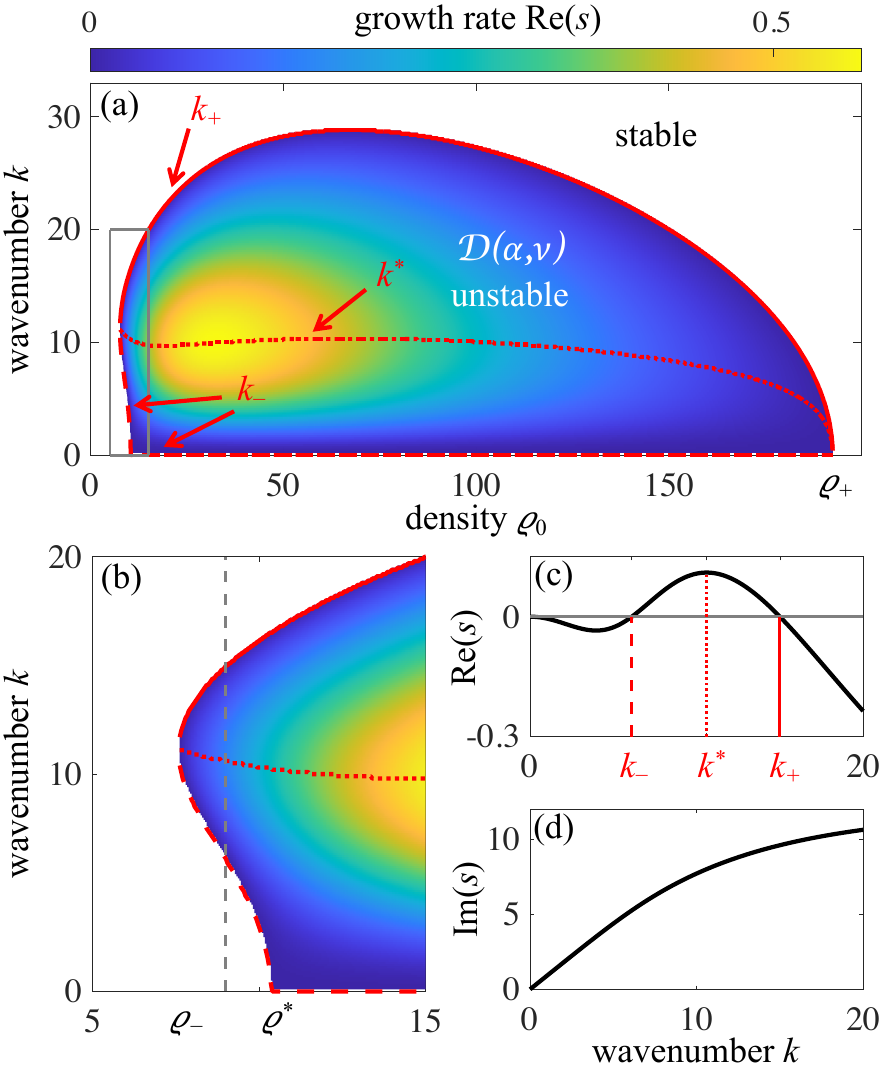}
    \end{center}
\caption{Results of the linear stability analysis of the uniform-density schooling state, as described in \S\ref{Sec:LinStab} of the Main Text and \S\ref{App:LinStab} of the Supplemental Material. The flapping frequency is $\alpha=4\pi$ and diffusivity $\nu = 0.05$. This figure is to be compared with Fig.~\ref{StabFig1} in the Main Text. The key difference is that $\alpha$ satisfies $\alpha > \alpha^*(\nu)$; this regime corresponds to scenario \#3 described in \S\ref{Sec:LinStab} of the Main Text, for which a finite-wavelength instability occurs at $\varrho=\varrho_-$ but a long-wave instability occurs at $\varrho=\varrho_+$. (a) Growth rate $\text{Re}(s)$ of the dominant eigenvalue, as a function of the school density $\varrho_0$ and perturbation wavenumber $k$. The stable region ($\text{Re}(s) < 0$) is indicated in white, and the unstable region $\mathcal{D}(\alpha,\nu)$ is colored. The wavenumbers $k_-$, $k^*$ and $k_+$ are indicated by the dashed, dotted and solid red lines, respectively. (b) Zoom-in of the region indicated by the gray box in panel (a), which highlights the finite-wavelength instability that occurs for densities $\varrho_0$ in the interval $(\varrho_-,\varrho^*)$. (c--d) Real (c) and imaginary (d) parts of the dominant eigenvalue $s(k)$ of the linear stability problem for $\varrho_0 = 9$, which is indicated by the vertical dashed line in panel (b). The fact that $k_->0$ in panel (c) indicates a finite-wavelength instability.}
\label{fig:FiniteAmp}
\end{figure}

Our strategy is to determine, for fixed $\alpha > 0$ and $\nu > 0$, the region $\mathcal{D}=\mathcal{D}(\alpha,\nu)$ in the $(\varrho_0,k)$--parameter plane for which the uniform-density state is unstable, $\text{Re}(s_1) > 0$. Examples of $\mathcal{D}$ are the colored regions in Fig.~\ref{StabFig1}c of the Main Text and Fig.~\ref{fig:FiniteAmp}a.

Our numerical computations indicate that, when the uniform-density state is unstable, it is so for wavenumbers in a single finite interval, $\text{Re}(s_1) > 0$ for $k\in (k_-,k_+)$ and $\text{Re}(s_1) < 0$ for $k > k_+$ (Fig.~\ref{StabFig1}a in the Main Text). While we do not have a formal proof of this observation, it is supported by numerical computations of the eigenvalues of the matrix $M(k)$. Our computations also indicate that, for $\alpha < \alpha_c(\nu)$, the uniform-density state is stable for all values of $\varrho_0$. An expression for $\alpha_c$ will be given in Eq.~\eqref{Crit1a}. For $\alpha > \alpha_c$, we observe that 
the boundary of $\mathcal{D}$ is a simple closed curve that may be decomposed into two parts: a single interval on the $\varrho_0$-axis ($k=0$), which we call $\mathcal{B}_0$, and a smooth curve $\mathcal{B}_1$. For $\alpha_c(\nu) < \alpha < \alpha^*(\nu)$, $\mathcal{B}_0$ is an interval $(\varrho_-,\varrho_+)$ and $\mathcal{B}_1$ is given by a single-valued function $k=k_+(\varrho_0)$ (Fig.~\ref{StabFig1}c in the Main Text), so the stability problem exhibits long-wave instabilities ($k_-=0$) for all $\varrho_0\in (\varrho_-,\varrho_+)$. For $\alpha > \alpha^*(\nu)$, $\mathcal{B}_1$ is given by a multi-valued function of $\varrho_0$ (Fig.~\ref{fig:FiniteAmp}), with finite wavelength ($k_->0$) instabilities in an interval $(\varrho_-,\varrho^*)$ and long-wave ones ($k_-=0$) in $(\varrho^*,\varrho_+)$. In Eqs.~\eqref{rhoM} and~\eqref{rhoP}, we obtain asymptotic expressions for the long-wave instability interval $(\varrho^{\text{L}}_-,\varrho^{\text{L}}_+)$ in the limit $\nu\rightarrow 0$, where $\varrho^{\text{L}}_+=\varrho_+$ and $\varrho^{\text{L}}_-=\varrho_-$ ($\varrho^{\text{L}}_-=\varrho^*$) for $\alpha < \alpha^*$ ($\alpha > \alpha^*$). We will also show in \S\ref{App:astar} that $\alpha^*\rightarrow 1+\sqrt{2}$ as $\nu\rightarrow 0$.

We proceed by finding an equation that describes $\mathcal{B}_1$. Defining $s=s_{\text{R}}+\rmi s_{\text{I}}$, $\mathcal{B}_1$ is defined by the solutions to $F(\rmi s_{\text{I}},k)=0$. The real and imaginary parts of this equation are
\begin{subequations}\label{BoundEq}
\begin{align}
a_0-b_1s_{\text{I}}-a_2s_{\text{I}}^2&=0,\label{BoundReal} \\
b_0+a_1s_{\text{I}}-b_2s_{\text{I}}^2-s_{\text{I}}^3&=0,\label{BoundImag}
\end{align}
\end{subequations}
where the real ($a_i$) and imaginary ($b_i$) parts of the functions $f_i(k)$ in Eq.~\eqref{fiFuncs} are
\begin{align}
a_0&=\nu k^2(1+\alpha^2),\quad a_1=-\frac{\varrho_0}{C_0}+2\nu k^2,\quad a_2=2+\varrho_0+\nu k^2,\nonumber \\
b_0&=-k(1+\alpha^2)(1+C_0),\quad b_1=-2k(1+C_0),\quad b_2=-k(1+C_0).\label{aibiCoeffs}
\end{align}
Multiplying Eq.~\eqref{BoundReal} by $s_{\text{I}}$, Eq.~\eqref{BoundImag} by $a_2$ and subtracting the two equations, we obtain
\begin{align}
-a_2b_0+(a_0-a_1a_2)s_{\text{I}}+(a_2b_2-b_1)s_{\text{I}}^2=0.\label{BoundInt}
\end{align}
Using Eq.~\eqref{BoundReal}, the $s_{\text{I}}^2$--term in Eq.~\eqref{BoundInt} can be eliminated:
\begin{align}
\left[(a_0-a_1a_2)a_2-b_1(a_2b_2-b_1)\right]s_{\text{I}}+(a_2b_1-b_1)a_0-a_2^2b_0=0.\label{BoundLin}
\end{align}
The positive solution $s_{\text{I}}$ of Eq.~\eqref{BoundReal} is
\begin{align}
s_{\text{I}}=k\frac{1+C_0+\sqrt{(1+C_0)^2+\nu(1+\alpha^2)(2+\varrho_0+\nu k^2)}}{2+\varrho_0+\nu k^2}.\label{sIsoln}
\end{align}
Substituting Eq.~\eqref{aibiCoeffs} and Eq.~\eqref{sIsoln} into Eq.~\eqref{BoundLin} and canceling out a factor of $k(1+C_0)$, we obtain
\begin{align}
&\phantom{=}G(k;\varrho_0,\alpha,\nu)\equiv\left\{-\frac{2(1+\alpha^2)^2k^2(\varrho_0+\nu k^2)}{(1+\alpha^2+\varrho_0)^2(2+\varrho_0+\nu k^2)}+(1+\alpha^2)\nu k^2-(2+\varrho_0+\nu k^2)(1+\alpha^2+\varrho_0+2\nu k^2)\right\}\nonumber \\
&\times\left(1+\sqrt{1+\nu\frac{(2+\varrho_0+\nu k^2)(1+\alpha^2+\varrho_0)^2}{1+\alpha^2}}\right)+(1+\alpha^2)\left[(2+\varrho_0)^2+\nu k^2(4+\varrho_0)\right]=0.\label{GBound}
\end{align}
For fixed $\alpha$ and $\nu$, the solution(s) to Eq.~\eqref{GBound} define $\mathcal{B}_1$. Due to the complexity of that equation, we proceed by studying its solutions in two limits: the long-wave limit $k\rightarrow 0$ (\S\ref{App:Long}) and the zero-diffusion limit $\nu=0$ (\S\ref{App:kplus}).

\subsection{Long-wave instabilities}\label{App:Long}

We now deduce the parameter regime in which long-wave instabilities ($k_-=0$) exist. To that end, we evaluate the limit of $G$ in Eq.~\eqref{GBound} as $k\rightarrow 0$:
\begin{align}
&\lim_{k\rightarrow 0}G(k;\varrho_0,\alpha,\nu)=(2+\varrho_0)\left(G_0-G_1\right),\quad\text{where}\quad G_0=1+\alpha^2(1+\varrho_0)\nonumber \\
&\text{and}\quad G_1=(1+\alpha^2+\varrho_0)\sqrt{1+\nu\frac{(2+\varrho_0)(1+\alpha^2+\varrho_0)^2}{1+\alpha^2}}.\label{G0Bound}
\end{align} 
Note that $G_0,G_1 > 0$. Multiplying and dividing by $G_0+G_1$, we obtain
\begin{align}
\lim_{k\rightarrow 0}G(k;\varrho_0,\alpha,\nu)=\frac{(2+\varrho_0)^2}{(1+\alpha^2)(G_0+G_1)}B,\quad
\text{where}\,\,\,\,B(\varrho_0,a,\nu)=\varrho_0(a-2)a^2-\nu(\varrho_0+a)^4\quad\text{and}\quad a=1+\alpha^2.\label{G0Bound2}
\end{align}
Note that $B$ is a quartic polynomial in $\varrho_0$, is negative both as $\varrho_0\rightarrow 0$ and $\varrho_0\rightarrow\infty$, and has a unique local maximum ($\partial B/\partial \varrho_0=0$) at $\varrho_{\text{max}}=\mu^{-1/3}-a$, where  $\mu=4\nu/[(a-2)a^2]$ and $B(\varrho_{\text{max}})=\nu\left(3\mu^{-1/3}-4a\right)/\mu$. Since $B(\varrho_{\text{max}}) > 0$ (and $\varrho_{\text{max}} > 0$) for $\mu < (3/4a)^3$, we conclude that if
\begin{align}
\alpha > \alpha_c\equiv\sqrt{\frac{2}{1-4\nu(4/3)^3}-1},\label{Crit1a}
\end{align}
there exist values $\varrho_0=\varrho_-^{\text{L}}$ and $\varrho_+^{\text{L}}$ with $\varrho_-^{\text{L}}<\varrho_+^{\text{L}}$ for which $\lim_{k\rightarrow 0}G(k;\varrho_0,\alpha,\nu)$ equals zero. We thus conclude that $\mathcal{B}_1$ intersects the horizontal axis ($k=0$) at the values $\varrho_0=\varrho_{\pm}^{\text{L}}$.

To show that $\mathcal{B}_0$ is the interval $(\varrho_-^{\text{L}},\varrho_+^{\text{L}})$, we need to show that $\text{Re}(s_1)>0$ in a neighborhood of $k=0$. 
Differentiating the equation $F(s_1(k),k)=0$ with respect to $k$ yields 
\begin{align}
\partial_sFs_1^{\prime}+\partial_kF=0,\label{Fk}
\end{align}
where primes denote differentiation with respect to $k$. We thus obtain
\begin{align}
s_1^{\prime}(0)=-\frac{\partial_kF(0,0)}{\partial_sF(0,0)}=-\frac{f_0^{\prime}(0)}{f_1(0)}=\rmi\left(\frac{1+\alpha^2}{1+\alpha^2+\varrho_0}\right)^2,
\end{align}
which agrees with Eq.~\eqref{EigLong}. The behavior of $\text{Re}(s_1)$ is deduced by differentiating Eq.~\eqref{Fk} with respect to $k$,
\begin{align}
\partial_sFs_1^{\prime\prime}+\partial_{ss}F(s_1^{\prime})^2+2\partial_{sk}Fs_1^{\prime}+\partial_{kk}F=0,\label{Fkk}
\end{align}
so
\begin{align}
s_1^{\prime\prime}(0)&=-\frac{\partial_{kk}F+2\partial_{sk}Fs_1^{\prime}+\partial_{ss}F(s_1^{\prime})^2}{\partial_sF}=-\frac{f_0^{\prime\prime}(0)+2f_1^{\prime}(0)s_1^{\prime}+2f_2(0)(s_1^{\prime})^2}{f_1(0)}=\frac{2a}{(a+\varrho_0)^5}B,\label{s2Deriv}
\end{align}
where all derivatives of $F$ are evaluated at $k=0$ and $s=0$ (see Eq.~\eqref{EigLong}), and $B$ was defined in Eq.~\eqref{G0Bound2}. We conclude that, for $\alpha > \alpha_c$ and $\varrho_0\in (\varrho_-^{\text{L}},\varrho_+^{\text{L}}$), $B>0$ so $\text{Re}[s_1^{\prime\prime}(0)]>0$, which in turn implies that $\text{Re}(s_1)>0$ in a neighborhood of $k=0$ because $\text{Re}[s_1(0)]=\text{Re}[s_1^{\prime}(0)]=0$ (Eq.~\eqref{EigLong}). That is, long-wave instabilities ($k_-=0$) exist provided that $\alpha > \alpha_c$ and $\varrho_0\in (\varrho_-^{\text{L}},\varrho_+^{\text{L}}$). In particular, a long-wave instability exists only if $\nu < (1/4)(3/4)^3\approx 0.106$. 
Note that Eq.~\eqref{Crit1a} is precisely the criterion given in Eq.~\eqref{Crit1} in the Main Text.
 
\subsubsection{Asymptotic expressions for $\varrho_{\pm}^{\text{L}}$ in the limit $\nu\rightarrow 0$}\label{App:qLqR}

For a fixed $\nu$ and $\alpha$ satisfying Eq.~\eqref{Crit1a}, the values $\varrho_{\pm}^{\text{L}}$ are given by the two real roots of $B$, for which analytical expressions are cumbersome. We proceed by deriving asymptotic expressions for $\varrho_{\pm}^{\text{L}}$ in the limit $\nu \rightarrow 0$. To that end, we write 
\begin{align}
B = (a-2)a^2D(\sigma,a,\mu)\quad\text{where}\quad D= \sigma-a- \frac{\mu\sigma^4}{4},\quad \sigma=a+\varrho_0\quad\text{and}\quad \mu = \frac{4\nu}{a^2(a-2)},
\end{align}
and approximate the roots of $D$ (a quartic polynomial in $\sigma$) in the limit $\mu\rightarrow 0$. Note that $D(\sigma,a,0)$ is a linear function in $\sigma$ with a single root at $\sigma=a$, so the other root of $D$ must go to infinity as $\mu\rightarrow 0$. 

Writing $\sigma=a+\mu\varsigma$, we find that
\begin{align}
D(a+\mu\varsigma,a,\mu)=-\frac{\mu a^4}{4}+(1-\mu a^3)\mu\varsigma+O(\mu^3),
\end{align}
which is $O(\mu^3)$ for $\varsigma=a^4/[4(1-\mu a^3)]$. We thus find that one root of $D$ may be approximated as 
\begin{align}
\sigma=a+\frac{\mu a^4}{4(1-\mu a^3)}+O(\mu^3)\quad\Rightarrow\quad \varrho_-^{\text{L}}=\frac{\nu(1+\alpha^2)^2}{\alpha^2-1-4\nu(1+\alpha^2)}+O(\nu^3).\label{rhoM}
\end{align}
The other real root of $D$, which goes to infinity as $\mu\rightarrow 0$, is obtained by writing $\sigma=(4/\mu)^{1/3}+\varsigma$. Indeed,
\begin{align}
D\left((4/\mu)^{1/3}+\varsigma,a,\mu\right)=-a-3\varsigma+O(\mu^{1/3}),
\end{align}
which is $O(\mu^{1/3})$ for $\varsigma=-a/3$. We thus find that the other root of $D$ may be approximated as 
\begin{align}
\sigma=(4/\mu)^{1/3}-a/3+O(\mu^{1/3})\quad\Rightarrow\quad \varrho_+^{\text{L}}=(\alpha^2+1)\left[\left(\frac{\alpha^2-1}{\nu(\alpha^2+1)}\right)^{1/3}-\frac{4}{3}\right]+O(\nu^{1/3}).\label{rhoP}
\end{align}
Equations~\eqref{rhoM} and~\eqref{rhoP} are written in Eq.~\eqref{rhoPM} of the Main Text.

\subsubsection{Asymptotic expression for $\alpha^*$ in the limit $\nu\rightarrow 0$}\label{App:astar}

We now examine the shape of the upper boundary $\mathcal{B}_1$ of $\mathcal{D}$ for $k\gtrsim 0$ and $\varrho_0\approx \varrho_{\pm}^{\text{L}}$. We restrict our attention to the regime $\nu \ll 1$, for which asymptotic expressions for $\varrho_{\pm}^{\text{L}}$ are available [see Eqs.~\eqref{rhoM} and~\eqref{rhoP}]. For fixed $\alpha$ and $\nu$, $\mathcal{B}_1$ is defined by $G(k,\varrho_0,\alpha,\nu)=0$, where $G$ is defined in Eq.~\eqref{GBound}. Differentiating with respect to $k$, we obtain $\partial_kG+\partial_{\varrho_0}G\varrho_0^{\prime}=0$. Since $\partial_kG=0$ at $k=0$, and
\begin{align}
\partial_{\varrho_0}G(0,\varrho_-^{\text{L}},\alpha,\nu)&=2(\alpha^2-1)+O(\nu),\quad
\partial_{\varrho_0}G(0,\varrho_+^{\text{L}},\alpha,\nu)=-\frac{3}{2\alpha^2}\left(\frac{\alpha^2-1}{\alpha^2+1}\right)^{1/3}(\alpha^2-1)(\alpha^2+1)^2\nu^{-1/3}+\dots,
\end{align}
(i.e. $\partial_{\varrho_0}G\neq 0$), we conclude that $\varrho_0^{\prime}=0$ at $\varrho_0=\varrho_{\pm}^{\text{L}}$ and $k=0$. That is, $\mathcal{B}_1$ intersects the $\varrho_0$--axis vertically (see Fig.~\ref{StabFig1}c and Fig.~\ref{fig:FiniteAmp}). Differentiating Eq.~\eqref{GBound} twice with respect to $k$, we obtain
\begin{align}
\partial_{kk}G+2\partial_{k\varrho_0}G\varrho_0^{\prime}+\partial_{\varrho_0\varrho_0}G(\varrho_0^{\prime})^2+\partial_{\varrho_0}G\varrho_0^{\prime\prime}=0.\label{G2k}
\end{align}
Evaluating Eq.~\eqref{G2k} at $k=0$ and using the fact that $\varrho_0^{\prime}=0$ there, we obtain the expression $\varrho_0^{\prime\prime}=-\partial_{kk}G/\partial_{\varrho_0}G$ at $k=0$. Using Eqs.~\eqref{rhoM} and~\eqref{rhoP}, we have the following expressions for $\partial_{kk}G$:
\begin{align}
\partial_{kk}G(0,\varrho_-^{\text{L}},\alpha,\nu)=4\nu\frac{\alpha^4-6\alpha^2+1}{\alpha^2-1}+O(\nu^2),\quad
\partial_{kk}G(0,\varrho_+^{\text{L}},\alpha,\nu)=-\frac{(\alpha^2+1)^{5/3}(5\alpha^4-2\alpha^2+1)}{\alpha^2(\alpha^2-1)^{2/3}}\nu^{2/3}+\dots.\label{GkkEq}
\end{align}
That is, $\partial_{kk}G < 0$ at $\varrho_0=\varrho_+^{\text{L}}$, so $\varrho_0^{\prime\prime}<0$ there; that is, $\mathcal{B}_1$ curves to the left at $\varrho_0=\varrho_+^{\text{L}}$ (see Fig.~\ref{StabFig1}c and Fig.~\ref{fig:FiniteAmp}). For $1 < \alpha <\alpha^*$, where $\alpha^*\approx 1+\sqrt{2}$ by Eq.~\eqref{GkkEq}, $\partial_{kk}G<0$ at $\varrho_0=\varrho_-^{\text{L}}$ so $\varrho_0^{\prime\prime} > 0$ there. That is, $\mathcal{B}_1$ curves to the right at $\varrho_0=\varrho_-^{\text{L}}$  for $1 < \alpha < \alpha^*$ (Fig.~\ref{StabFig1}c). However, it curves to left at $\varrho_0=\varrho_-^{\text{L}}$ for $\alpha > \alpha^*$ (Fig.~\ref{fig:FiniteAmp}).

We note that $\alpha^*$ is defined as the value for which $\mathcal{B}_1$ has an inflection point at $k=0$, $\varrho_0=\varrho_-^{\text{L}}$, specifically, 
\begin{align}
\partial_{kk}G(0,\varrho_-^{\text{L}}(\alpha^*,\nu),\alpha^*,\nu)=0,\label{astar}
\end{align}
where $\varrho_0=\varrho_-^{\text{L}}$ is the smaller root of $B$ in Eq.~\eqref{G0Bound2}. The numerical solutions $\alpha^*(\nu)$ of Eq.~\eqref{astar} are shown in Table~\ref{tab:astar}, and they approach $1+\sqrt{2}$ in the limit $\nu\rightarrow 0$, as expected.

\begin{table}
\begin{center}
\begin{tabular}{|c||c|c|c|c|c|c|}
\hline   $\nu$ & 0.05  & 0.01 & 0.005 & 0.001 & 0.0005 & 0.0001 \\ \hline
   $\alpha^*$     & 3.1542 & 2.5042 & 2.4571 &  2.4222 & 2.4181 & 2.4147 \\[1mm] 
\hline
\end{tabular}
\end{center}
\caption{Numerical computations  of the solutions $\alpha^*$ of Eq.~\eqref{astar} for various values of $\nu$.}
\label{tab:astar}
\end{table}

\subsection{The zero-diffusion regime $\nu=0$}\label{App:kplus}

The results in \S\ref{App:Instab} may be simplified considerably in the zero-diffusion regime, $\nu=0$. The upper boundary $\mathcal{B}_1$ of the unstable region $\mathcal{D}$ is determined by evaluating Eq.~\eqref{GBound} for $\nu=0$:
\begin{align}
G(k;\varrho_0,\alpha,0)=\varrho_0(2+\varrho_0)\left[\alpha^2-1-\left(\frac{2(1+\alpha^2)}{(2+\varrho_0)(1+\alpha^2+\varrho_0)}\right)^2k^2\right]=0.\label{Gnu0}
\end{align}
We conclude that the uniform-density state is unstable for any $\varrho_0 > 0$ and $\alpha > 1$, because Eq.~\eqref{Gnu0} has the positive solution
\begin{align}
k_+=\frac{(1+\alpha^2+\varrho_0)(2+\varrho_0)}{2(1+\alpha^2)}\sqrt{\alpha^2-1},\label{k_plus2}
\end{align}
which is the expression given in Eq.~\eqref{k_plus} in the Main Text. We note that Eq.~\eqref{k_plus2} furnishes a good approximation for $k_+$ for $0 < \nu \ll 1$ provided $\varrho_0$ is not too close to the stability boundaries $\varrho_{\pm}$. Specifically, if $\alpha>\alpha_c$ and $\varrho_0$ are fixed, Eq.~\eqref{k_plus2} approximates the largest unstable wavenumber $k_+$ as $\nu\rightarrow 0$. However, while $k_+$ increases quadratically with $\varrho_0$ for $\nu=0$ and assumes the value $k_+=\sqrt{\alpha^2-1}$ for $\varrho_0=0$ (Eq.~\eqref{k_plus2}), our numerical computations show that $k_+$ has a non-monotonic dependence on $\varrho_0$ for $\nu > 0$ (Fig.~\ref{StabFig1}c in the Main Text and Fig.~\ref{fig:FiniteAmp}a). Specifically, $k_+$ increases as $\varrho_0$ is increased from $\varrho_-$, but eventually decreases and approaches zero as $\varrho_0\uparrow \varrho_+$.

{For $\nu=0$, the short-wave ($k\rightarrow \infty$) behavior of the eigenvalues $s_1$ and $s_2$ is the same as that given in Eq.~\eqref{EigShort}. To determine the behavior of $s_3$, we posit that $s_3\sim \mathring{s}_{\text{R}}+\rmi k\mathring{s}_{\text{I}}$ as $k\rightarrow \infty$. To determine $\mathring{s}_{\text{R}}$ and $\mathring{s}_{\text{I}}$, we find the dominant contributions to $F(s_3;k)$ as $k\rightarrow\infty$:
\begin{align}
s_3^3&\sim -\rmi \mathring{s}_{\text{I}}^3k^3-3\mathring{s}_{\text{R}}\mathring{s}_{\text{I}}^2k^2+\dots,\quad f_2(k)s_3^2\sim \rmi \frac{1+\alpha^2}{1+\alpha^2+\varrho_0}\mathring{s}_{\text{I}}^2k^3+k^2\left(2\frac{1+\alpha^2}{1+\alpha^2+\varrho_0}\mathring{s}_{\text{R}}\mathring{s}_{\text{I}}-(2+\varrho_0)\mathring{s}_{\text{I}}^2\right)+\dots,\nonumber \\
f_1(k)s_3&\sim 2\frac{1+\alpha^2}{1+\alpha^2+\varrho_0}\mathring{s}_{\text{I}}k^2+\dots.
\end{align}
Setting $F(s_3;k)=0$, we thus obtain
\begin{align}
\mathring{s}_{\text{I}}=\frac{1+\alpha^2}{1+\alpha^2+\varrho_0}\quad\text{and}\quad \mathring{s}_{\text{R}}=-\varrho_0,
\quad\text{so}\quad s_3\sim -\varrho_0+\rmi\frac{1+\alpha^2}{1+\alpha^2+\varrho_0}k\quad\text{as}\quad k\rightarrow\infty.
\end{align}
}

\begin{figure}[ht]
   \begin{center}
    \includegraphics[width=0.6\textwidth]{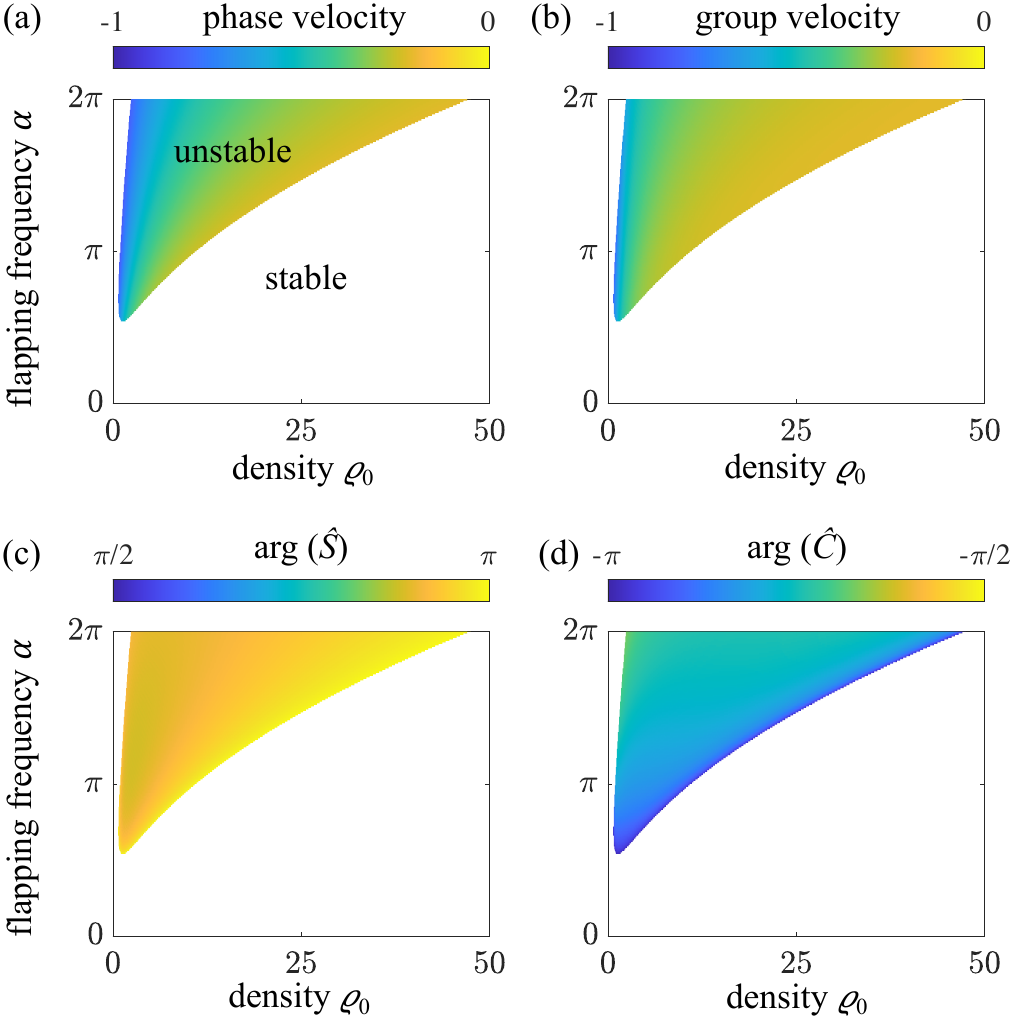}
    \end{center}
\caption{Features of the most unstable mode with wavenumber $k^*$ according to the linear stability analysis of the uniform-density schooling state, as described in \S\ref{Sec:LinStab} of the Main Text and \S\ref{App:LinStab} of the Supplemental Material. The diffusivity is fixed at $\nu = 0.05$. (a) Dependence of the phase velocity $-\text{Im}(s)/k^*$ on the density $\varrho_0$ and flapping frequency $\alpha$. (b) Dependence of the group velocity $-(\mathrm{d}/\mathrm{d}k)\text{Im}(s)$, evaluated at $k=k^*$, on the density $\varrho_0$ and flapping frequency $\alpha$. (c--d) Phase angle of the components $\hat{S}$ (panel c) and $\hat{C}$ (panel d) of the eigenvector $(\hat{\varrho},\hat{C},\hat{S})$ of $M(k^*)$, which indicates the manner in which the mode destabilizes. We take the first component of the eigenvector to be unity, $\hat{\varrho}=1$, without loss of generality. Note that the colorbar ranges are different in panels (c) and (d), in order to accentuate the variations in the displayed quantities.}
\label{fig:MiscStab}
\end{figure}

\section{Numerical simulations of the continuum model}\label{Sec:RungeKutta}

In Figs.~\ref{Fig:Snapshots},~\ref{Fig:TWaveStab} and~\ref{Fig:Compact} of the Main Text, we presented the results of numerical simulations of the dimensionless continuum model in Eq.~\eqref{PDE1} of the Main Text. The equations were solved with periodic boundary conditions on the interval $[-L\pi,L\pi]$ via a pseudospectral method. Specifically, we write $\hat{\varrho}(k,t)=\mathcal{F}[\varrho(x,t)]$, $\hat{C}(k,t)=\mathcal{F}[C(x,t)]$ and $\hat{S}(k,t)=\mathcal{F}[S(x,t)]$, $\mathcal{F}$ being the Fourier transform in $x$, and obtain
\begin{align}
\diff{}{t}\hat{\bs{\xi}}=-\mathcal{L}\hat{\bs{\xi}}+\hat{\bs{N}},\quad\text{where}\quad\hat{\bs{\xi}}=\begin{pmatrix} \hat{\varrho} \\ \hat{C} \\ \hat{S}\end{pmatrix},\quad \mathcal{L}=\begin{pmatrix} -\rmi k+\nu k^2 & 0 & 0 \\ 1+\rmi\nu k & 1 & \alpha \\ 0 & -\alpha & 1\end{pmatrix},\quad \text{and}\quad \bs{N}=\begin{pmatrix} \pdiff{}{x}(\varrho C) \\ -\varrho C \\ 0\end{pmatrix}.\label{MatEqn1a}
\end{align}
The linear part of Eq.~\eqref{MatEqn1a} is treated exactly using an integrating factor: 
\begin{align}
\diff{}{t}\left(\rme^{\mathcal{L}t}\hat{\bs{\xi}}_k\right)=\rme^{\mathcal{L}t}\hat{\bs{N}}.
\label{MatEqn1b}
\end{align}
We solve Eq.~\eqref{MatEqn1b} using a fourth-order Runge-Kutta time-stepping scheme with a fixed time step $\Delta t$:
\begin{align}
\hat{\bs{\xi}}^{n+1} = \rme^{-\mathcal{L}\Delta t}\hat{\bs{\xi}}^{n} + \frac{\Delta t}{6}\left[\rme^{-\mathcal{L}\Delta t}\bs{f}_1+2\rme^{-\mathcal{L}\Delta t/2}(\bs{f}_2+\bs{f}_3)+\bs{f}_4\right]
\end{align}
where $\hat{\bs{\xi}}^{n}(k)=\hat{\bs{\xi}}(k,n\Delta t)$ and 
\begin{align}
\bs{f}_1&=\hat{\bs{N}}\left(\hat{\bs{\xi}}^n\right),\quad
\bs{f}_2  =\hat{\bs{N}}\left(\rme^{-\mathcal{L}\Delta t/2}\left(\hat{\bs{\xi}}^n+\frac{\Delta t}{2}\bs{f}_1\right)\right),\quad
\bs{f}_3 = \hat{\bs{N}}\left(\rme^{-\mathcal{L}\Delta t/2}\hat{\bs{\xi}}^n + \frac{\Delta t}{2}\bs{f}_2\right)\nonumber \\
\text{and}\quad\bs{f}_4 &= \hat{\bs{N}}\left(\rme^{-\mathcal{L}\Delta t}\hat{\bs{\xi}}^n+\Delta t\rme^{-\mathcal{L}\Delta t/2}\bs{f}_3\right).\label{fterms}
\end{align}
The matrix exponential has the form
\begin{align}
\rme^{-\mathcal{L}\Delta t} = \begin{pmatrix} \rme^{(\rmi k-\nu k^2)\Delta t} & 0 & 0 \\ \varphi\left[(1-\nu k^2+\rmi k)\left(\rme^{-\Delta t}\cos\alpha\Delta t-\rme^{(\rmi k-\nu k^2)\Delta t}\right)-\alpha\rme^{-\Delta t}\sin\alpha\Delta t\right] & \rme^{-\Delta t}\cos\alpha\Delta t & -\rme^{-\Delta t}\sin\alpha\Delta t \\ \varphi\left[\alpha\rme^{-\Delta t}\cos\alpha\Delta t-\alpha\rme^{(\rmi k-\nu k^2)\Delta t}+(1-\nu k^2+\rmi k)\rme^{-\Delta t}\sin\alpha\Delta t\right] & \rme^{-\Delta t}\sin \alpha\Delta t & \rme^{-\Delta t}\cos\alpha\Delta t\end{pmatrix}
\end{align}
where
\begin{align}
\varphi= \frac{1+\rmi \nu k}{\left[1-\nu k^2+\rmi (k+\alpha)\right]\left[1-\nu k^2+\rmi (k-\alpha)\right]}.
\end{align}
{For the simulations in this paper, we used 512 spatial grid points and a time step $\Delta t = 2^{-8}$, unless otherwise stated, which we found to be sufficient to resolve the structure and dynamics of the traveling wave solutions. An analogous method is used to numerically solve the quasistatic model in Eq.~\eqref{PDEQS} of the Main Text, which is easier to solve because it consists of a single unknown $\varrho_{\text{q}}(x,t)$.}

\section{Numerical method for finding traveling wave solutions}\label{App:TWave}


We here present a numerical method for finding traveling wave solutions in a periodic domain to the continuum model in Eq.~\eqref{PDE1} of the Main Text. These solutions satisfy the ODEs in Eq.~\eqref{TWave1}, the first of which may be integrated once to obtain
\begin{align}
\nu \varrho^{\prime}&=c\varrho-(1+C)\varrho-d,
\end{align}
where $d$ is a constant of integration. Equation~\eqref{TWave1} is thus equivalent to the system
\begin{align}
\nu \varrho^{\prime}&=c\varrho-(1+C)\varrho-d,\quad
cC^{\prime}=-c\varrho-C-\alpha S+d,\quad
cS^{\prime}=-S+\alpha C.\label{TWaveEq2a}
\end{align}
Equation~\eqref{TWaveEq2a} has a fixed point $(\varrho_0,C_0,S_0)$, which satisfies
\begin{align}
C_0 = \frac{d-c\varrho_0}{1+\alpha^2},\quad S_0 = \alpha C_0,
\end{align}
and the following quadratic equation for $\varrho_0$:
\begin{align}
c\varrho_0^2-\left[d+(1+\alpha^2)(1-c)\right]\varrho_0-d(1+\alpha^2)=0.\label{RPoly1}
\end{align}
Equation~\eqref{RPoly1} may be solved to express $d$ in terms of $\varrho_0$ and $c$:
\begin{align}
d = \frac{c\varrho_0^2-(1+\alpha^2)(1-c)\varrho_0}{1+\alpha^2+\varrho_0}\quad\Rightarrow\quad C_0 = -\frac{\varrho_0}{1+\alpha^2+\varrho_0},
\end{align}
which is identical to the expression derived in Eq.~\eqref{UniformState} of the Main Text.

\begin{figure} 
   \begin{center}
     \includegraphics[width=.8\textwidth]{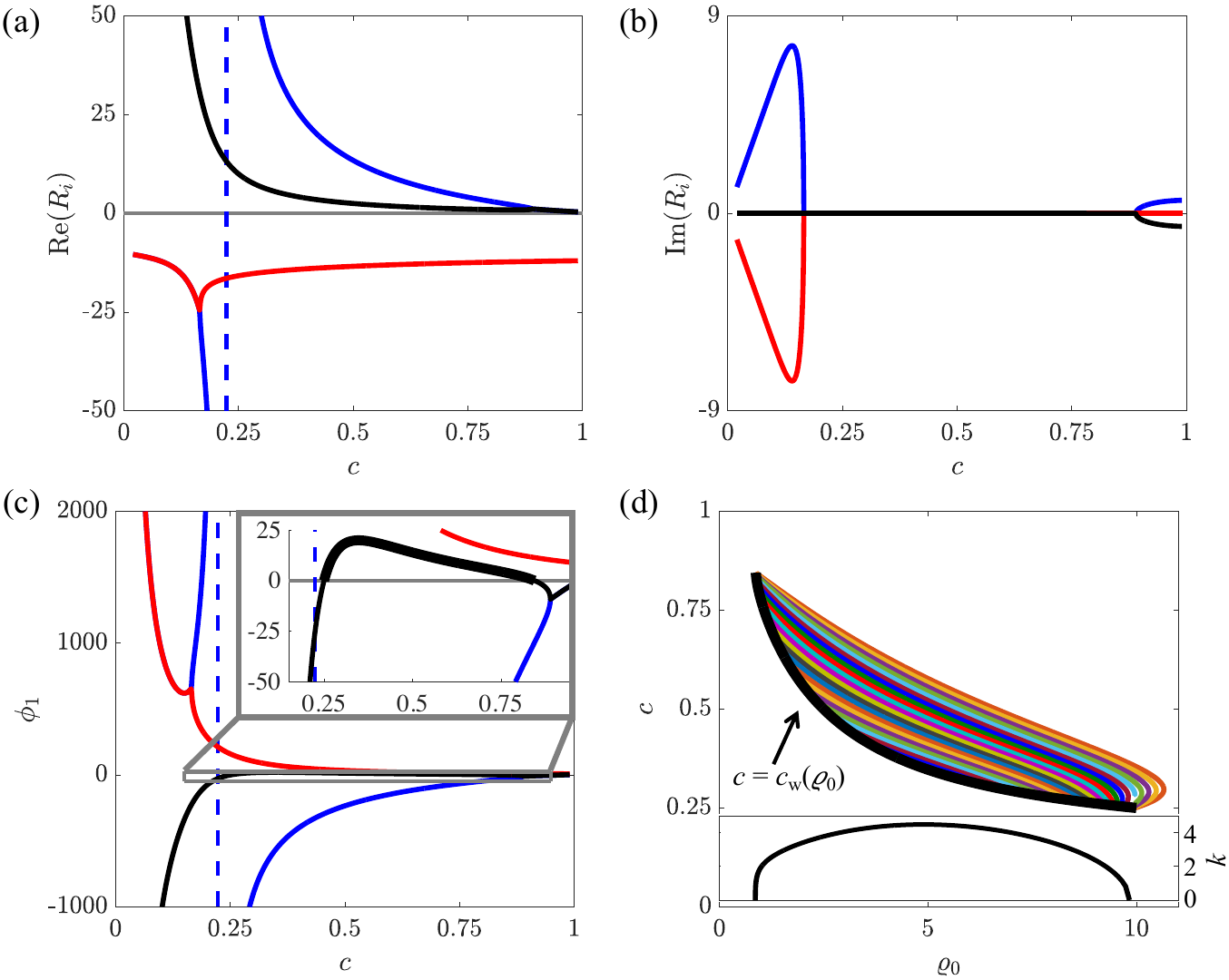}
      \end{center}
\caption{Illustration of the numerical method for finding traveling wave solutions, as described in \S\ref{App:TWave}, for fixed values of the parameters $\alpha=3$ and $\nu=0.05$. (a--b) Plots of the real (panel a) and imaginary (panel b) parts of the three roots $\varrho_0=R_i(c)$ ($1\leq i\leq 3$) of the cubic polynomial $\tilde{Q}(\varrho_0,c)$ defined in Eq.~\eqref{Qtilde}, where $R_1$, $R_2$ and $R_3$ are shown in black, blue and red, respectively. The dashed line in panel (a) indicates that $\text{Re}(R_2(c))\rightarrow\pm\infty$ as $c\rightarrow\sqrt{\nu}^{\pm}$, since the leading coefficient $q_3$ of $\tilde{Q}$ vanishes (see Eq.~\eqref{Qtilde}). (c) Plot of the functions $\phi_1(R_i(c),c)$ for each of the three roots $\varrho_0=R_i(c)$. The inset contains a magnification of the region shown, the thicker black line indicating the portion of the curve for which $\phi_1(R_1(c),c) > 0$. (d) The colored curves show the numerically computed traveling wave solution branches, reproduced from Fig.~\ref{fig:twave}b in the Main Text. The black curve shows a portion of the black curve from panel (a), restricted to the interval on which $\phi_1(R_1(c),c) > 0$ (see panel (c) inset) and with the axes inverted. The inset shows the dependence of the wavenumber $k = \sqrt{\phi_1(\varrho_0,R_1^{-1}(\varrho_0))}$ on the density $\varrho_0$.
}
\label{Fig:TWaveCalc}
\end{figure}

We proceed by linearizing Eq.~\eqref{TWaveEq2a} around the fixed point $(\varrho_0,C_0,S_0)$, and so substitute the expressions $\varrho=\varrho_0+\epsilon\tilde{\varrho}$, $C=C_0+\epsilon\tilde{C}$ and $S=S_0+\epsilon\tilde{S}$ into Eq.~\eqref{TWaveEq2a} and retain terms at leading order in $\epsilon$. We thus obtain $\bs{\zeta}^{\prime}=\tilde{M}\bs{\zeta}$, where $\bs{\zeta}=(\varrho,C,S)$ and
\begin{align}
\tilde{M}= \begin{pmatrix} (c-C_0-1)/\nu & -\varrho_0/\nu & 0 \\ -1 & -1/c & -\alpha/c \\ 0 & \alpha/c & -1/c\end{pmatrix}.\label{LinMat1}
\end{align}
Small-amplitude traveling waves bifurcate from the curve $c= c_{\text{w}}(\varrho_0)$ in the $(\varrho_0,c)$--plane on which $\tilde{M}$ possesses a pair of purely imaginary eigenvalues $\pm \rmi k$, where $k\in\mathbb{R}$. The characteristic polynomial of $\tilde{M}$ is
\begin{align}
 \Phi(s;\varrho_0,c)\equiv\det(\tilde{M}-sI)=-\phi_0-\phi_1s-\phi_2s^2-s^3,
 \end{align}
 where
 \begin{align}
 \phi_0(\varrho_0,c) &= \frac{(\alpha^2+1)^2-c(1+\alpha^2+\varrho_0)^2}{c^2\nu(1+\alpha^2+\varrho_0)},\quad \phi_1(\varrho_0,c)=\frac{(1+\alpha^2+\varrho_0)\left[\nu(1+\alpha^2)-c^2(2+\varrho_0)\right]+2c(1+\alpha^2)}{c^2\nu(1+\alpha^2+\varrho_0)}\nonumber \\
 \text{and}\quad\phi_2(\varrho_0,c) &=\frac{(1+\alpha^2+\varrho_0)(2\nu-c^2)+c(1+\alpha^2)}{c\nu(1+\alpha^2+\varrho_0)}.\label{phiExp1}
 \end{align}
We note that $\Phi(s;\varrho_0,c)$ is related to the characteristic polynomial $F(s;k)$ of the matrix $M$ defined in Eq.~\eqref{MMat} by $F(\mathrm{i}kc;k)=-\Phi(\mathrm{i}k;\varrho_0,c)\mathrm{i}kc^2\nu$. The curve $c=c_{\text{w}}(\varrho_0)$ corresponds to the values of $c$ for which $\Phi(s;\varrho_0,c) = -(s^2+k^2)(s-\hat{s})$, where $\hat{s}\in\mathbb{R}$. This requirement implies that $\phi_1 > 0$ and that
  \begin{align}
  Q(\varrho_0,c)\equiv \phi_1\phi_2-\phi_0=0.\label{TCond}
  \end{align}
The roots of $Q$ may readily be found by writing
 \begin{align}
 Q(\varrho_0,c) = \frac{\tilde{Q}(\varrho_0,c)}{c^3\nu^2(1+\alpha^2+\varrho_0)^2},
 \end{align}
 where $\tilde{Q}$ is a cubic polynomial in $\varrho_0$:
 \begin{align}
 \tilde{Q}(\varrho_0,c) &= q_0(c)+q_1(c)\varrho_0+q_2(c)\varrho_0^2+q_3(c)\varrho_0^3,\nonumber \\
 \text{where}\quad q_0(c) &= 2(1+\alpha^2)^2\left[c^4-2c^3+c^2(1-2\nu)+2c\nu+\nu^2(1+\alpha^2)\right],\nonumber \\
q_1(c) &= (1+\alpha^2)\left[(5+\alpha^2)(c^4-c^3)-c^2\nu(9+\alpha^2)+4c\nu+4(1+\alpha^2)\nu^2\right],\nonumber \\
 q_2(c) &= 2c^4(2+\alpha^2)-c^3(1+\alpha^2)-2c^2\nu(3+\alpha^2)+2(1+\alpha^2)\nu^2\quad\text{and}\quad q_3(c) = c^2(c^2-\nu).\label{Qtilde}
 \end{align}
 The functions $Q$ and $\tilde{Q}$ share the same three roots $\varrho_0=R_i(c)$ for $1\leq i\leq 3$, which are computed numerically for fixed values of the parameters $\alpha$ and $\nu$ and shown in Fig.~\ref{Fig:TWaveCalc}a--b. Figure~\ref{Fig:TWaveCalc}c shows a plot of $\phi_1(R_i(c),c)$, and the inset shows that there is an interval of values of $c$ on which $\phi_1(R_1(c),c) > 0$. The endpoints of this interval satisfy $\phi_0(R_1(c),c)=0$ and $\phi_1(R_1(c),c)=0$. The first condition implies that
 \begin{align}
 1+\alpha^2+R_1=\frac{1+\alpha^2}{\sqrt{c}}.
 \end{align}
We substitute this into the condition $\phi_1(R_1,c)=0$ and, upon eliminating $R_1$, find that the endpoints of the interval are the solutions to the equation
\begin{align}
c^{3/2}(1-\sqrt{c}) = \nu\frac{\alpha^2+1}{\alpha^2-1}.\label{ccrit1}
\end{align}
The maximum value of the left-hand side of~\eqref{ccrit1} is $\nu^*\equiv 3^3/4^4$, so Eq.~\eqref{ccrit1} has two solutions provided that
\begin{align}
\alpha > \sqrt{\frac{\nu^*+\nu}{\nu^*-\nu}},
\end{align}
which is equivalent to the condition given in Eq.~\eqref{Crit1} of the Main Text. 

We define the curve $c_{\text{w}}$ as $c_{\text{w}}(\varrho_0)=R_1^{-1}(\varrho_0)$, restricted to values of $c$ for which $\phi_1(R_1(c),c) >0$, as shown in Fig.~\ref{Fig:TWaveCalc}d. The inset shows the wavenumber $k = \sqrt{\phi_1(\varrho_0,R_1^{-1}(\varrho_0))}$. We thus employ the following algorithm to construct traveling wave solutions on a periodic domain of length $2\pi L$. Consider a branch of solutions for which $\varrho$ has $n$ minima, and let $R_n^{+}$ and $R_n^{-}$ be the larger (smaller) values of $\varrho_0$ for which $k = n/L$ (Fig.~\ref{Fig:TWaveCalc}d, inset). We discretize the ODEs in Eq.~\eqref{TWave1} of the Main Text using the fast Fourier transform {with 512 points} and numerically solve the resulting nonlinear algebraic equations using MATLAB's $\tt fsolve$ package, which employs a trust-region-dogleg algorithm. We use 
\begin{align}
\varrho(\xi) = R_n^++A\sin (n\pi \xi/L),\quad C(\xi) = -\frac{\varrho(\xi)}{1+\alpha^2+\varrho(\xi)},\quad S(\xi)=\alpha C(\xi)\quad\text{and}\quad c=c_{\text{w}}(R_n^+)
\end{align}
as the initial guess for the solution with amplitude $A = 0.01$ that bifurcates from $\varrho_0 = R_n^+$. The initial guesses for $C$ and $S$ are motivated by the corresponding expressions for a uniform-density state, as given in Eq.~\eqref{UniformState} in the Main Text. We proceed through numerical continuation in the parameter $A$ until $A$ achieves its maximum value, after which we switch to the mean density $\bar{\varrho}=(2\pi L)^{-1}\int_{-\pi L}^{\pi L}\varrho(\xi)\,\mathrm{d}\xi$ as the continuation parameter. We decrement $\bar{\varrho}$ progressively until the endpoint $\bar{\varrho}=R_n^-$.

The colored curves in Fig.~\ref{Fig:TWaveCalc}d indicate the numerically computed traveling wave solutions, reproduced from Fig.~\ref{fig:twave}b in the Main Text. By construction, each traveling wave branch begins and ends on the curve $c=c_{\text{w}}(\varrho_0)$, and the two corresponding values of $k$ are equal. 


\section{Simulations of ``finite schools" with compactly supported initial data}\label{App:Compact}

{In \S\ref{Sec:TWave} of the Main Text, we identified traveling wave solutions to the continuum model Eq.~\eqref{PDE1} on a periodic domain. To assess whether traveling wave-like behavior would emerge in a finite school, in \S\ref{Sec:Compact} of the Main Text we conducted simulations of Eq.~\eqref{PDE1} with initial data that is compactly supported and $C^{\infty}$:
\begin{align}
\varrho(x,0)&=\left(\varrho_0+\tilde{\varrho}(x)\right)\begin{cases} 1 &\text{if }|x| < \kappa_1, \\ \psi\left(\frac{\kappa_2-|x|}{\kappa_2-\kappa_1}\right)&\text{if }\kappa_1\leq |x|\leq \kappa_2, \\ 0 &\text{if }|x|>\kappa_2, \end{cases}\quad\text{where}\quad \psi(x)=\frac{\rme^{-1/x}}{\rme^{-1/x}+\rme^{-1/(1-x)}},\nonumber \\
C(x,0) &= -\frac{\varrho(x,0)}{1+\alpha^2+\varrho(x,0)},\quad S(x,0) = \alpha C(x,0),\label{CompactIC}
\end{align}
and  $\tilde{\varrho}(x)$ is a small perturbation. That is, for $\tilde{\varrho}(x)=0$ the swimmers would be equidistributed within the interval $[-\kappa_1,\kappa_1]$, where $\kappa_1$ is much smaller than the length $2\pi L$ of the simulation domain. The swimmer density decreases to zero exponentially in the interval $\kappa_1 < |x| < \kappa_2$. The choices for $C(x,0)$ and $S(x,0)$ are obtained from the uniform-density solution given in Eq.~\eqref{UniformState}.}

{Simulations of the continuum model with initial data given by Eq.~\eqref{CompactIC} are shown in Fig.~\ref{Fig:Compact} in the Main Text and Supplemental Movie 2. The parameters are $\kappa_1=5\pi$, $\kappa_2=5\pi+64$, and the simulation domain length is $2\pi L$ where $L = 320$. The simulation was conducted with $2^{14}=16384$ spatial grid points and a time step $\Delta t = 2^{-12}$, the higher resolution being required due to the large size of the domain. }

\subsection{Quasistatic model with $\nu=0$}\label{App:QSNuZero}

\begin{figure}[ht]
   \begin{center}
    \includegraphics[width=1\textwidth]{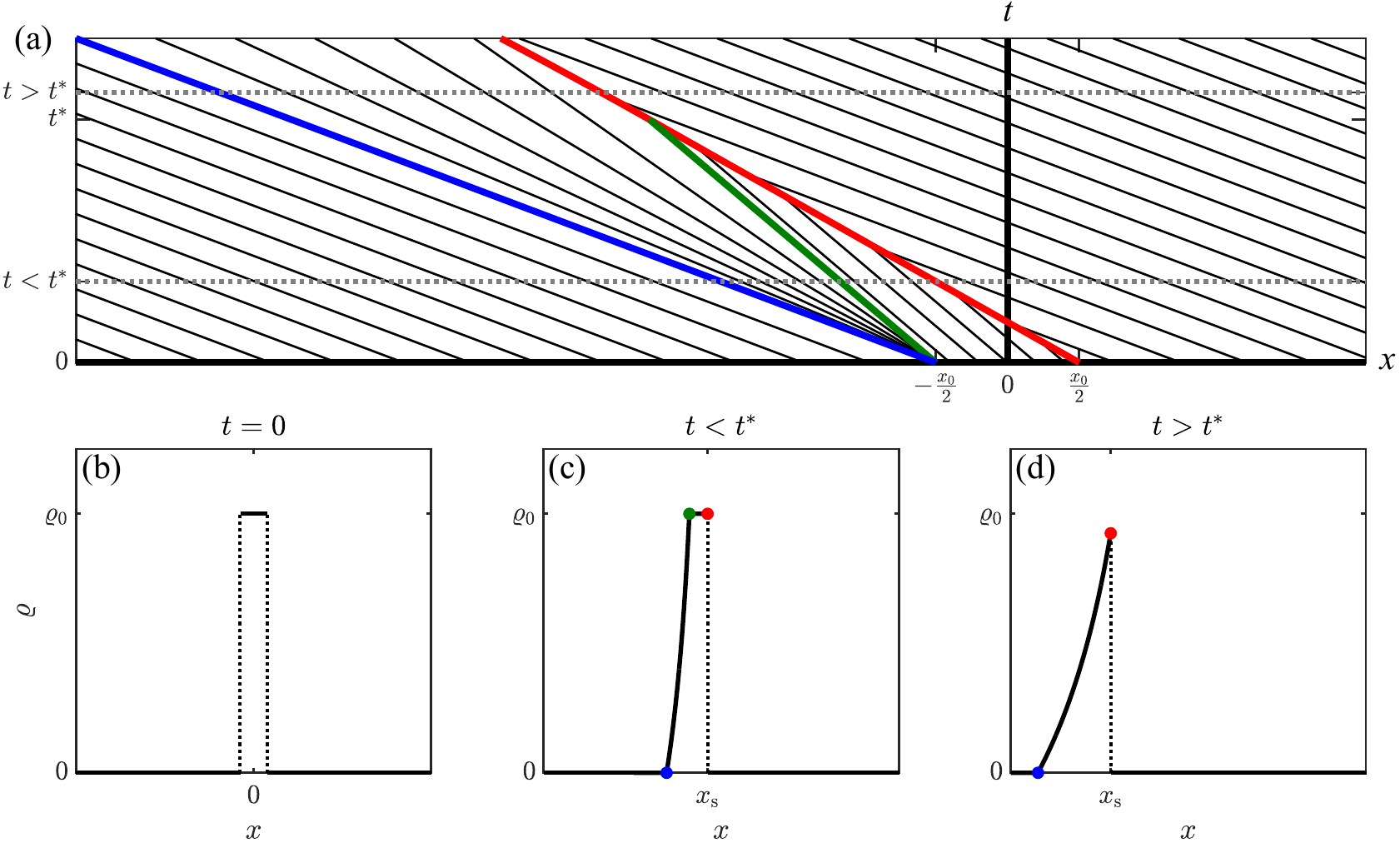}
    \end{center}
\caption{{The method of characteristics is used to solve the quasistatic model~\eqref{PDEQS} without diffusion ($\nu=0$) with rectangular initial data. (a) Characteristics in the $(x,t)$ plane. The blue curve is the upstream edge of the rarefaction fan, $x=-x_0/2-t$. The green curve is the downstream edge $x_{\mathrm{f}}(t)$ of the rarefaction fan, given by Eq.~\eqref{xf}. The red curve is the shock position $x_{\mathrm{s}}(t)$, given by Eq.~\eqref{xs1} for $t <t^*$ and Eq.~\eqref{xs2} for $t > t^*$. (b) Initial data, given by Eq.~\eqref{RectIC}. (c) Solution $\varrho(x,t)$ for $t < t^*$, as given by Eq.~\eqref{rhotLts}. (d) Solution $\varrho(x,t)$ for $t > t^*$, as given by Eq.~\eqref{rhotGts}.}}
\label{fig:QSNuZero}
\end{figure}

{We now solve the quasistatic model in Eq.~\eqref{PDEQS} of the Main Text without diffusion ($\nu=0$) and with the rectangular initial data
\begin{align}
\varrho(x,0)=\begin{cases} 0 &\text{if }|x| > x_0/2, \\ \varrho_0 &\text{if }|x| < x_0/2.\end{cases}\label{RectIC}
\end{align}
The constant $x_0$ is chosen so that the total mass is equal to that in Eq.~\eqref{CompactIC} with $\tilde{\rho}=0$:
\begin{align}
x_0 &= 2\kappa_1+2\int_{\kappa_1}^{\kappa_2}\rmd x\,\psi\left(\frac{\kappa_2-x}{\kappa_2-\kappa_1}\right)=2\kappa_1+2(\kappa_2-\kappa_1)\int_0^1\rmd x\,\psi\left(1-x\right)=\kappa_1+\kappa_2,
\end{align}
where in the last step we use the facts that $\psi(x)+\psi(1-x)=1$ and $\int_0^1\rmd x\,\psi(x)=\int_0^1\rmd x\,\psi(1-x)$. The initial data is shown in Fig.~\ref{fig:QSNuZero}b.}

{The solution can be derived using the method of characteristics. Defining the flux
\begin{align}
q(\varrho)=\varrho U_\text{q}(\varrho),
\end{align}
where the velocity $U_{\text{q}}$ is given in Eq.~\eqref{PDEQS}, the characteristic equations are
\begin{align}
\diff{x}{t}=q^{\prime}(\varrho)=-\left(\frac{1+\alpha^2}{1+\alpha^2+\varrho}\right)^2,\quad \diff{\varrho}{t}=0.\label{CharEq}
\end{align}
The characteristics are thus lines in the $(x,t)$-plane, as shown in Supplementary Figure~\ref{fig:QSNuZero}a. The downstream edge of the rarefaction fan (green curve) is given by
\begin{align}
x_{\mathrm{f}}(t)=-\frac{x_0}{2}-\left(\frac{1+\alpha^2}{1+\alpha^2+\varrho_0}\right)^2t.\label{xf}
\end{align}
The shock trajectory $x_{\mathrm{s}}(t)$ (red curve) is determined by the Rankine-Hugoniot jump condition:
\begin{align}
\diff{x_{\mathrm{s}}}{t}&=\frac{[q]}{[\varrho]}=-\frac{1+\alpha^2}{1+\alpha^2+\varrho_0}, \quad x_{\mathrm{s}}(0)=\frac{x_0}{2},
\end{align}
where $[f]=f(x_{\mathrm{s}}^+)-f(x_{\mathrm{s}}^-)$ denotes the jump in the quantity $f$. The solution to this ODE is
\begin{align}
x_{\mathrm{s}}(t)&=\frac{x_0}{2}-\frac{1+\alpha^2}{1+\alpha^2+\varrho_0}t\quad\text{for}\quad t < t^*,\label{xs1}
\end{align}
where $t^*$ is the time at which the shock and downstream edge of the rarefaction intersect: 
\begin{align}
x_{\mathrm{f}}(t^*)=x_{\mathrm{s}}(t^*)\quad\Rightarrow \quad t^*=x_0\frac{(1+\alpha^2+\varrho_0)^2}{(1+\alpha^2)\varrho_0}.\label{tstar}
\end{align}
The solution is thus
\begin{align}
\varrho(x,t)=\begin{cases} 0 &\text{if }x < -t-\frac{x_0}{2}, \\ (1+\alpha^2)\left(\sqrt{-\frac{t}{x+x_0/2}}-1\right) &\text{if }-t-\frac{x_0}{2} < x < x_{\mathrm{f}}(t), \\ \varrho_0 &\text{if }x_{\mathrm{f}}(t) < x < x_{\mathrm{s}}(t), \\ 0 &\text{if }x > x_{\mathrm{s}}(t),\end{cases}\quad\text{for}\quad t < t^*,\label{rhotLts}
\end{align}
which is depicted in Fig.~\ref{fig:QSNuZero}c. The shock trajectory for $t > t^*$ is given by
\begin{align}
\diff{x_{\mathrm{s}}}{t}&=\frac{[q]}{[\varrho]}=-\sqrt{\frac{-x_{\mathrm{s}}}{t}},\quad x_{\mathrm{s}}(t^*)=-\frac{x_0}{2}-\frac{1+\alpha^2}{\varrho_0}x_0\nonumber \\
\Rightarrow\quad x_{\mathrm{s}}(t)&=-\frac{x_0}{2}-\left(\sqrt{t}+\sqrt{\frac{(1+\alpha^2)x_0}{\varrho_0}}-\sqrt{t^*}\right)^2\quad\text{for}\quad t > t^*.\label{xs2}
\end{align}
The solution is thus
\begin{align}
\varrho(x,t)=\begin{cases} 0 &\text{if }x < -t-\frac{x_0}{2}, \\ (1+\alpha^2)\left(\sqrt{-\frac{t}{x+x_0/2}}-1\right) &\text{if }-t-\frac{x_0}{2} < x < x_{\mathrm{s}}(t), \\ 0 &\text{if }x > x_{\mathrm{s}}(t),\end{cases}\quad\text{for}\quad t > t^*,\label{rhotGts}
\end{align}
which is depicted in Fig.~\ref{fig:QSNuZero}d.}

\newpage
\section*{Supplemental Movies}

\noindent {\bf Supplemental Movie 1:} Numerical simulation of the continuum PDE model [Eq.~\eqref{PDE1}] shows how small perturbations to a uniform-density schooling state amplify into a self-sustaining traveling wave. The parameters are listed in the caption of Fig.~\ref{Fig:Snapshots}. The panels show, from left to right, the density $\varrho(x,t)$, mean velocity $U_0(x,t)=-1-C(x,t)$ and field $S(x,t)$.  The red dots in the first panel depict ``Lagrangian" particles that evolve with the mean velocity $U_0$, which serves to illustrate the evolving density. \\

\noindent {\bf Supplemental Movie 2:} {Numerical simulation of the continuum PDE model [Eq.~\eqref{PDE1}] with compactly supported initial data shows how a ``finite school" goes unstable via internal traveling waves. The parameters $\alpha$ and $\nu$ are the same as those in Supplemental Movie 1. The simulation parameters are listed in Supplemental Material \S\ref{App:Compact}. The top (bottom) row shows the solution for $\varrho_0 = 3.5$ ($\varrho_0 = 0.7$), which is predicted to be unstable (stable) by the linear theory presented in \S\ref{Sec:LinStab} of the Main Text. The panels show, from left to right, the density $\varrho(x,t)$, mean velocity $U_0(x,t)=-1-C(x,t)$ and field $S(x,t)$. The red dashed curves show the quasistatic solution, obtained by solving Eq.~\eqref{PDEQS}. The green curves show the (discontinuous) solution to the quasistatic model without diffusion ($\nu=0$), as given in Supplemental Material \S\ref{App:QSNuZero}. The solution corresponding to $\varrho_0 = 3.5$ (top row) is shown in Fig.~\ref{Fig:Compact}. Note that the videos show a portion of the full solution in order to make the oscillations visible.}

\end{document}